\documentclass[a4paper,USenglish]{lipics}

\usepackage{microtype} % Slightly tweak font spacing for aesthetics

\usepackage{tablefootnote}
\usepackage{algorithm}% http://ctan.org/pkg/algorithms
\usepackage{algpseudocode}% http://ctan.org/pkg/algorithmicx

\newcommand{\var}[1]{\textrm{\texttt{#1}}}% variable

\usepackage{color}
\usepackage{cleveref}
\usepackage{paralist} % Used for the compactitem environment which makes bullet points with less space between them
\usepackage{booktabs}

\newcommand{\bigoh}{\mathcal{O}}

\usepackage{thmtools, thm-restate}
\usepackage{xcolor}
\usepackage{tikz}
\usepackage[normalem]{ulem}

\renewcommand{\qed}{\hfill $\Box$}

\newcommand{\dft}{\textsc{DFT-Tree}}
\newcommand{\dfts}{\textsc{DFT-Trees}}
\title{Dynamic subtree queries revisited:\newline the Depth First Tour Tree}%\footnote{This work was partially supported by someone.}}
\titlerunning{Dynamic subtree queries revisited: the Depth First Tour Tree} %optional, in case that the title is too long; the running title should fit into the top page column

\author[1]{Gabriele Farina}
\author[2]{Luigi Laura}
\affil[1]{Polytechnic University of Milan, Italy\\
  \texttt{gabriele2.farina@mail.polimi.it}}
\affil[2]{``Sapienza'' Universiy of Rome, Italy\\
  \texttt{laura@dis.uniroma1.it}}
\authorrunning{G. Farina and L. Laura} %mandatory. First: Use abbreviated first/middle names. Second (only in severe cases): Use first author plus 'et. al.'

\Copyright{Gabriele Farina and Luigi Laura}%mandatory, please use full first names. LIPIcs license is "CC-BY";  http://creativecommons.org/licenses/by/3.0/

% mandatory: Please choose ACM 1998 classifications from http://www.acm.org/about/class/ccs98-html . E.g., cite as "F.1.1 Models of Computation". 
% mandatory: Please provide 1-5 keywords
\serieslogo{}%please provide filename (without suffix)
\volumeinfo%(easychair interface)
  {Billy Editor and Bill Editors}% editors
  {2}% number of editors: 1, 2, ....
  {Conference title on which this volume is based on}% event
  {1}% volume
  {1}% issue
  {1}% starting page number
\EventShortName{}
\DOI{10.4230/LIPIcs.xxx.yyy.p}% to be completed by the volume editor
\newcommand{\runinsec}[1]{\noindent\textbf{\textsf{#1}}\quad}

\begin{document}
%----------------------------------------------------------------------------------------
%	ABSTRACT
%----------------------------------------------------------------------------------------
\maketitle

\begin{abstract}
\noindent In the \emph{dynamic tree problem} the goal is the maintenance of an arbitrary $n$-vertex forest, where the trees are subject to joining and splitting by, respectively, adding and removing edges. Depending on the application, information can be associated to nodes or edges (or both), and queries might require to combine values in path or (sub)trees.

In this paper we present a novel data structure, called the \emph{Depth First Tour Tree}, based on a linearization of a DFS visit of the tree. Despite the simplicity of the approach, similar to the ET-Trees (based on a Euler Tour), our data structure is able to answer queries related to both paths and (sub)trees. In particular, focusing on subtree computations, we show how to customize the data structure in order to answer queries for three distinct applications: impact of the removal of an articulation point from a graph, betweenness centrality and closeness centrality of a dynamic tree. 
\end{abstract}

%----------------------------------------------------------------------------------------
%	ARTICLE CONTENTS
%----------------------------------------------------------------------------------------

\section{Introduction}

In the \emph{dynamic tree problem} the goal is the maintenance of an arbitrary $n$-vertex forest, where the trees are subject to joining and splitting by, respectively, adding and removing edges. Depending on the application, information can be associated to nodes or edges (or both), and queries might require to combine values in path or (sub)trees.

\begin{figure}[t!]
\centering
\includegraphics[trim=40 120 40 0, clip, width = 0.44\linewidth]{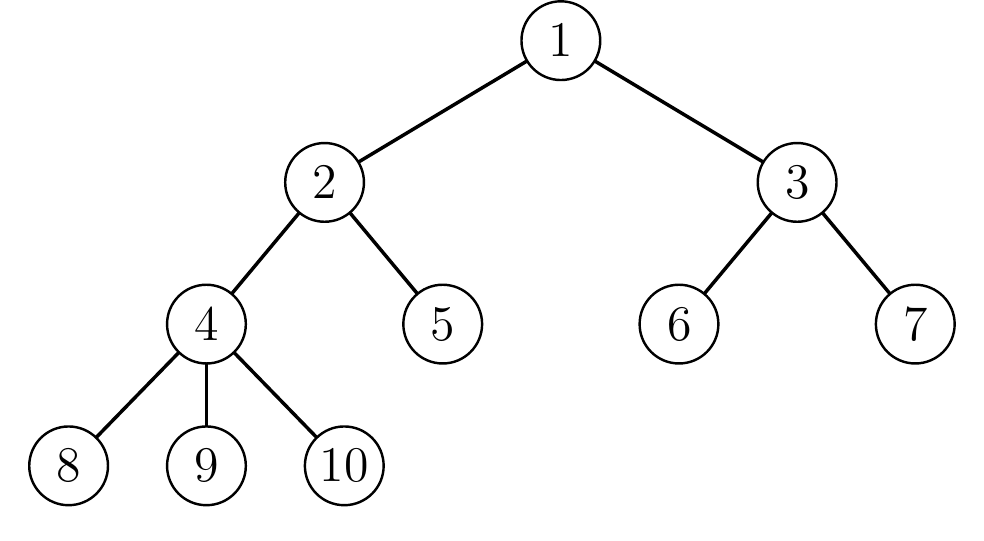}\hfill
\raisebox{.7cm}{\includegraphics[trim=0 0 0 160, clip, width = 0.53\linewidth]{asy/dft_vs_et.pdf}}
\caption{\label{fig:dftvseuler}An example of Euler Tour, Depth First Tour and \emph{parenthetical sequence} of a tree (introduced in Section~\ref{sec:dft}).}
\end{figure}

The dynamic tree problem has several applications, ranging from network flows~\cite{AMO93,GGT91,ST85,Tar97}, one of the original motivations, to other graph algorithms including connectivity~\cite{HK99}, biconnectivity~\cite{Fre83}, and minimum spanning trees~\cite{HK99,Fre85}, and other combinatorial problems~\cite{KMT03,Lan00}. 
With such a wealth of applications, it is not surprising the fact that there are several approaches to solve (at least partially) the dynamic tree problem using $\bigoh(\log n)$ time per operation: ST-trees~\cite{ST83,ST85}, ET-trees~\cite{HK99,Tar97}, topology trees~\cite{Fre85,Fre97a,Fre97b}, top trees~\cite{AHLT97,AHLT05,TW05}, RC-trees~\cite{ABV04,ABV05}, and Mergeable Trees~\cite{GKSTW11} that build up on the ST-tree and, as the name suggests, support also the \emph{merge} operation. All these approaches map a generic tree into a balanced one, and can be divided into three main categories: \emph{path decomposition} (ST-trees, Mergeable Trees), \emph{tree contraction} (topology trees, top trees, RC-trees), and \emph{linearization} (ET-trees);  refer to the dissertation of Werneck~\cite{Wer06} and the experimental comparison of Tarjan and Werneck~\cite{TW09} for a more complete picture about techniques and applications.
\\[4mm]
\runinsec{Approach.} In this paper we present a novel data structure, called the \emph{Depth First Tour Tree} (\dft), to solve the dynamic tree problem; the \dft, as the ET-Tree, is based on a linearization: as the name suggests, we linearize the tree following a DFS visit of it (see Figure~\ref{fig:dftvseuler}, where is shown for comparison also the Euler Tour). The main consequence of this approach is that the whole subtree of a node is stored contiguously, thus allowing us fast operations on the subtree, as we will detail in the rest of the paper.  As we can see from Figure~\ref{fig:dftvseuler}, for example, the subtree of node $4$ is contiguous in the \dft, whilst node $4$ itself appears twice in its own subtree in the corresponding ET-Tree. \dft\ data structure can be easily implemented on top of any Balanced Binary Search Tree (BBST), such as Splay Trees~\cite{ST85} and Red-Black Trees~\cite{CLRS09}. 

The idea of linearizing the tree according to its DFS visit and maintaining the linearization in an efficient data structure is not new in the literature. Indeed, the very idea was exploited in other works, most notably \cite{JR12,MR01,navarro2014fully}, in the context of succinct trees. However, given the additional constraint of succinctness, the focus of these works is inherently different, and the set of supported queries is weaker and less oriented to data-processing operations.

The \dft\ supports all the operations shown in Table~\ref{tbl:DFTcomplexity}, that are divided in three groups: i) structural operations, i.e. the ones that alter the structure of the tree, ii) structural queries, and iii) operations related to the values stored in the vertices; as we can see, it supports all the traditional dynamic tree operations together with others, such as \textsc{lca} and \textsc{condense}, that are not completely standard and, thus, not supported by all the data structures;  \textsc{condense}, in particular, allows to use the \dft\ to implement the Block Forest structure, following the exact algorithm of Westbrook and Tarjan~\cite{WT92}. 

Furthermore, the \dft\ supports three non standard \emph{generic} operations, to be customized depending on the applications, that are:
\begin{compactitem}
\item \textsc{combine}$(v)$, that aggregates values in the path between vertex $v$ and the root of the tree;
\item \textsc{reduce-children}$(v)$, that aggregates values of the children of $v$;
\item \textsc{reduce-child-subtrees}$(v)$, that aggregates values in the subtrees rooted in the children of $v$. 
\end{compactitem}
These \emph{generic} functions are, probably, the most interesting aspect of \dfts.
\\[3mm]
\runinsec{Contribution.}
% Easy to understand, analyze and implement
% Oriented to subtree-operations (ST trees are more path-oriented)
% Extremely flexible
% We give a unified framework
% We give applications
We propose a novel data structure, combining the simplicity of the Euler Tour trees with the expressiveness of the Depth First visit of a tree. We believe that the contribution of our approach is twofold:
\begin{compactitem}
	\item the resulting data structure is simple, using only elementary concepts, and thus is easy to understand, analyze and implement;
	\item we give a \emph{unified framework} for treating a vast class of data aggregation tasks on subtrees.
\end{compactitem}
While our data structure is able to support basic operations on paths, it is primarily designed to aggregate data on subtrees, an operation which is usually non-trivial with other data structures.

Unlike ST-trees, topology trees and RC-trees, \dfts\ do not require the underlying forest to have vertices with bounded (constant) degree in order to efficiently cope with subtree queries. Degree restrictions can be avoided by \emph{ternarizing} the input forest but, as observed in \cite{Wer08}, ``this introduces a host of special cases'' and complicates the data structure. In the special case of ST-trees, some work has been done \cite{Radzik98implementationof} to support queries on subtrees {for a restricted set of operations} (for example, giving the minimum element of a given subtree) without the need for ternarization, but the resulting data structure is still very complicated, both to analyze and implement. The same task can be performed extremely easily with \dfts.

Furthermore, \dfts\ can naturally aggregate on all the children subtrees of a node $v$ in parallel without having to pay a cost proportional to the degree of $v$ itself: for example, as we will see, given a node $v$ it takes $\bigoh(\log n)$, independently from the degree of $v$, to answer the child of $v$ whose subtree is the largest. This is an interesting feature that distinguishes our data structure, and can be useful for practical problems, as we will demonstrate in the final sections of this paper.

The extreme flexibility of use of the structure comes at the cost of its structural rigidity. In particular, while all other structural operations require logarithmic time in the forest size, the \textsc{evert} operation requires a cost proportional to the depth of the node being everted. However, when either the number of eversions is small compared to the total number of queries performed, or the costs of the eversion is amortized, the cost of \textsc{evert} can be regarded as being $\bigoh(\log n)$ like all the other structural operations. This is the case in all the applications we present.
\\[3mm]
\runinsec{Applications.}  In order to explain the versatility of the approach, we show how to customize the above functions for three distinct applications, based on subtree computations:
\begin{compactitem}
\item Given a streaming graph, for which we maintain all the biconnected properties using the mentioned approach of Westbrook and Tarjan, we can also compute the \emph{impact} of an articulation point $u$, introduced in the context of the Autonomous Systems (AS) graph, as a measure of the resiliency of the network. The impact of $u$ is defined as the number of vertices that gets disconnected from the main connected components after the removal of $u$. This application requires the determination of the subtree of a node  having maximum size.
\item The \emph{betweenness centrality} of a vertex $v$ in a tree. This requires to count the sum of the squares of the sizes inside subtrees. 
\item The \emph{closeness centrality} of a vertex $v$ in a tree. This requires the sum of the distances to every node in the subtree and in the tree above $v$. 
\end{compactitem}

In each of the above applications, the query on a vertex  can be executed in time $\bigoh ( \log n)$ for an $n$-vertices dynamic forest.

This paper is organized as follows: we conclude this section by recalling few preliminary notions. In Section~\ref{sec:dft} we describe the main ideas of the \dft, detailing the operations related to subtrees and paths in Section~\ref{sec:operations}. In Section~\ref{sec:applications} we show how to customize the generic operations of the \dfts\ in order to support the applications listed above.  Due to space constraints, we omit the proofs and low-level details such as the extensions of the operations to (edge-)weighted forests. More details about the implementation of the \dfts\ operations can be found, together with the pseudocode, in the Appendix. 

%Section~\ref{sec:conclusion} addresses concluding comments and remarks.

\newpage
% % % % % % % % % % % % % % % % % % % %
% Complexity table                    %
% % % % % % % % % % % % % % % % % % % %
\begin{table}[t!]

			\centering
			\begin{tabular}{lcl}
				\toprule
					Operation & Complexity & Description\\
				\midrule
					
					\textsc{link}$(u, v)$  & $\bigoh(\log n)$ & Makes the root of the tree containing vertex $v$\\
					& & a child of vertex $u$.\\
					\textsc{cut}$(v)$ & $\bigoh(\log n)$ & Deletes the edge connecting $v$ to its parent,\\
					& & splitting the tree. If $v$ is the root of the tree,\\
					& & nothing happens.\\
					\textsc{condense}$(v)$ & $\bigoh(\log n)$ & Deletes vertex $v$; its children  become children\\
					& & of the parent of $v$. If vertex $v$ is the root, the\\
					& & number of connected components of the forest\\
					& & increases by $d-1$, with $d$ being the degree of $v$.\\
					\textsc{erase}$(v)$ & $\bigoh(\log n)$ & Deletes vertex $v$ and all its adjacent edges.\\
					\textsc{evert}$(v)$ & $\bigoh(d \log n)$\tablefootnote{Where $d$ is the depth of the node involved. We note that the \textsc{evert} operation is slow in the worst case, but it is possible to amortize it by always everting the smallest tree.} & Re-roots the tree containing vertex $v$ at vertex $v$.\\
				\midrule
					\textsc{root}$(v)$ & $\bigoh(\log n)$ & Returns the root of the tree containing node $v$.\\
					\textsc{same-tree}$(u,v)$ & $\bigoh(\log n)$ & Tests if  nodes $u$ and  $v$ belong to the same tree.\\ 
					\textsc{is-descendant}$(u,v)$ & $\bigoh(\log n)$ & Answers whether  node $u$ is a descendant of $v$.\\
					\textsc{parent}$(v)$ & $\bigoh(\log n)$ & Returns the parent of node $v$.\\
					\textsc{ancestor}$(v, k)$ & $\bigoh(\log n)$ & Returns the ancestor of node $v$ at depth $d_v - k$,\\
					& &  where $d_v$ represents the depth of $v$, if existent.\\
					\textsc{lca}$(u, v)$ & $\bigoh(\log n)$ & Returns the lowest common ancestor of nodes $u$\\ 
					& &  and $v$ (if they belong to the same tree).\\
					\textsc{degree}$(v)$ & $\bigoh(\log n)$ & Returns the degree of node $v$.\\
					\textsc{list-children}$(v)$ & $\bigoh(\delta \log n)$\tablefootnote{Where $\delta$ is the degree of the node passed as argument to \textsc{degree}.} & Returns a list containing the children of vertex $v$.\\		
				\midrule
					\textsc{change-val}$(v, x)$ & $\bigoh(\log n)$ & Assigns \textrm{val}$(v) = x$.\\
					\textsc{reduce-children} & $\bigoh(\log n)$\tablefootnote{Assuming that the operations (denoted with $\oplus$ and $\otimes$) in \textsc{reduce-children}, \textsc{reduce-child-subtrees} and \textsc{combine} take constant time when called with two nodes.} & \emph{See description in the text, Section~\ref{sec:operations}.}\\
					\textsc{reduce-child-subtrees} & $\bigoh(\log n)$\footnotemark[\value{footnote}] & \emph{See description in the text, Section~\ref{sec:operations}.}\\
					\textsc{combine} & $\bigoh(\log n)$\footnotemark[\value{footnote}] & \emph{See description in the text, Section~\ref{sec:operations}.}\\
				\bottomrule
			\end{tabular}
			\vspace{1mm}
			\caption{\dft\ operations on an $n$ vertex forest. The complexity values reported are amortized complexity if we implement the \dft\ with Splay Trees~\cite{ST85} and worst-case complexity if we use Red-Black Trees~\cite{CLRS09}.\label{tbl:DFTcomplexity}}

		\end{table}

\section{Preliminaries} We assume the reader is familiar with basic concepts of graph theory (see, e.g., \cite{Die10}). We recall that, in an undirected graph $G$, a \emph{connected component} is a maximal set of vertices $V'\subseteq V$ such that, given $u,v\in V'$, there is at least one path between $u$ and $v$ in $G$; an \emph{articulation point} is a vertex $v\in V$ such that its removal from the graph $G$ increases the number of connected components of $G$; similarly a \emph{bridge} is an edge $e\in E$ such that its removal from the graph $G$ increases the number of connected components of $G$. A \emph{biconnected component} is a maximal set of vertices $V''\subseteq V$ such that after the removal of any $v\in V''$, the remaining graph $V''/v$ is connected. Following~\cite{AFL12trac}, the \emph{impact} of an articulation point is the number of vertices that get disconnected from the largest connected component when $v$ is removed from the graph.

There are several measures of centrality of vertices in a network. In this work we refer to the betweenness centrality and closeness centrality. The \emph{betweenness centrality}, originally defined in~\cite{freeman1977}, is defined as follows: $bc(u)=\sum_{s\neq t\neq v}\frac{\sigma_{st}(u)}{\sigma_{st}}$ where $\sigma_{st}(u)$ is the number of shortest paths between $s$ and $t$ that pass through $u$, and $\sigma_{st}$ is the total number of shortest paths.
The \emph{closeness centrality}, proposed by Bavelas in 1950~\cite{Bav50}, is the reciprocal of the \emph{farness} of a vertex, where the farness is the sum of all the distances to the other vertices in the graph. %\footnote{Note that, according to the original definition, if the graph is disconnected the closeness centrality of each vertex is $0$; in this paper, when we mention the closeness centrality of a vertex, we refer to the distances in the tree the vertex belongs. If the tree is a single vertex, or the farness is $0$, the closeness centrality is not defined.}.

%------------------------------------------------

\section{Depth First Tour Trees}
\label{sec:dft}

In this section we describe the main idea of the \dfts, which builds up on the \emph{Depth First Visit} of the tree and its linearization into an array; for the sake of the exposition we will populate this array with (opening and closing) parentheses that will be denoted as the \emph{parenthetical sequence} of the tree. %\footnote{Note that, despite the \emph{parenthetical sequence} we describe can be seen as the \emph{balanced parentheses} proposed by Munro and Raman ~\cite{MR01} in the context of succinct representation of trees, we do not investigate in this paper the applications of our approach in the field of succinct data structures; we refer the interested reader to the already mentioned paper of Munro and Raman and, for succinct dynamic tree maintenance,  to the works of Joannou and Raman~\cite{JR12}: here the focus is on a compact representation of the tree structure, providing basic navigations and updates, without queries that allow to aggregate values on paths or (sub)trees.}.
The other key ingredient of the \dfts\ is a summary defined over the parenthetical sequence: in the underlying BBST the  node corresponding to vertex $v$ is augmented with both the information about $v$ and the summary of its subtree (in the BBST).
The depth first visit of a tree is constructed by recursively visiting nodes in a depth-first fashion. When a node is entered for the first time, it is appended to the back of depth first tour, along with a tag indicating it was a newly-opened node (called an \emph{open-node}); when all its children have been visited, we push back the node again before returning the call, this time with tag indicating this is a fully explored node (called a \emph{close-node}). Since every node is appended to the list exactly twice, the size of the depth first tour of a tree of size $n$ is $2n$. 

\begin{figure}[t]
\centering
\includegraphics[trim=0 100 0 0, clip, width = 0.51\linewidth]{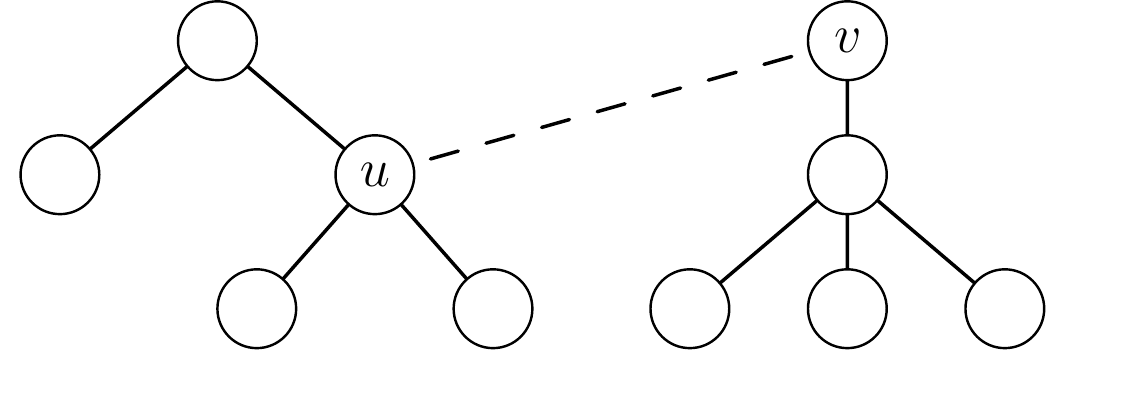}\hfill\raisebox{.6cm}{
\includegraphics[trim=0 0 0 110, clip, width = 0.47\linewidth]{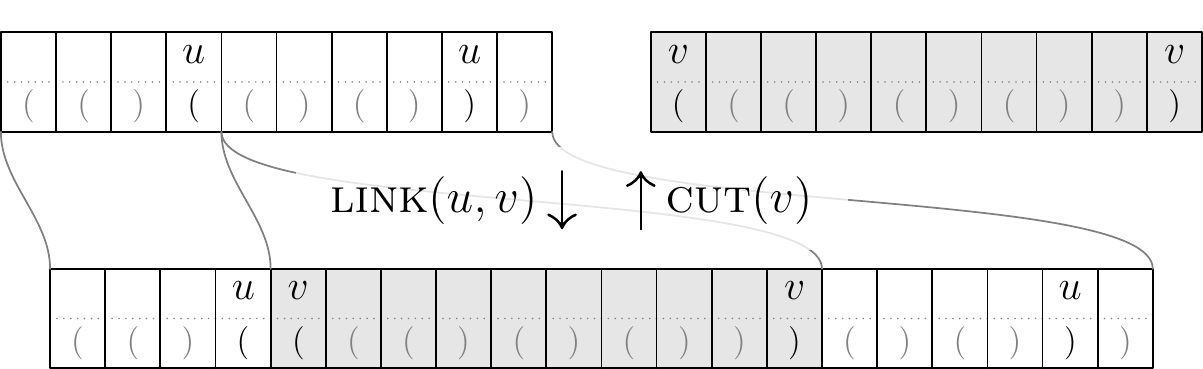}}\\[-4mm]
\textcolor{black!20!white}{\hrule}\vspace{4mm}
\includegraphics[width = 1\linewidth]{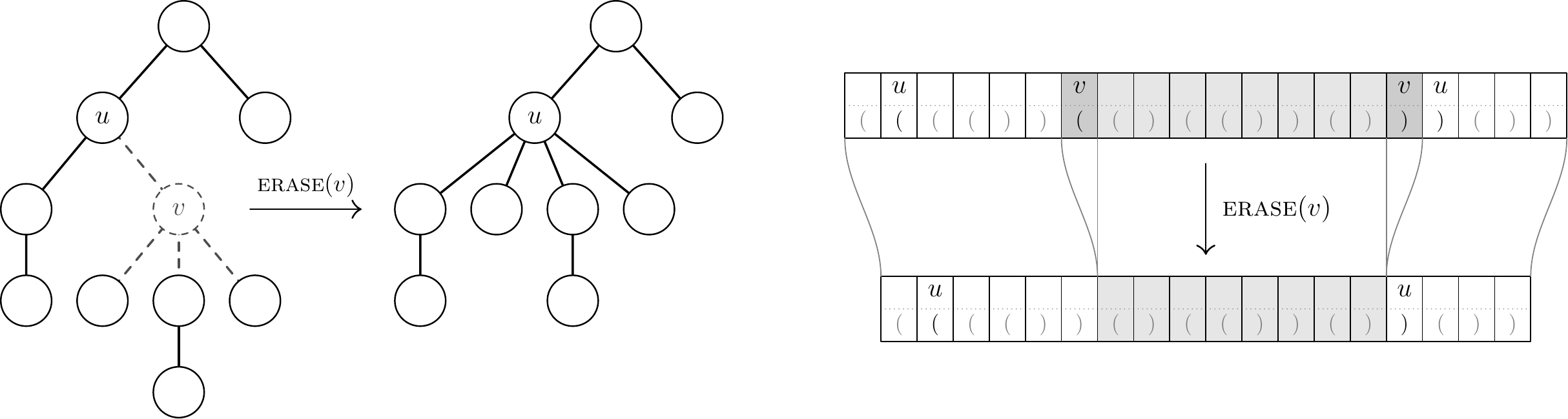}
\caption{Effects of the \textsc{link}, \textsc{cut} and \textsc{condense} operations.\label{fig:linkcut}}
\end{figure}

Figure \ref{fig:dftvseuler} shows the depth first tour of an example tree of size 10, together with its linearization: an array that contains its parenthetical sequence; the Euler Tour of the same tree is shown for comparison: note that in an Euler Tour a node can appear several time; the size of an Euler Tour is $1+2m=2n-1$, since an Euler Tour begins with a node and then, for each edge of the tree, both its endpoints are added exactly once, when entering the node. In Figure \ref{fig:linkcut} we can see the effects of the \textsc{link}, \textsc{cut} and \textsc{condense} operations on the tree and the corresponding parenthetical sequence.	

		\begin{definition}[depth of a parenthesis]
		We define the \emph{depth} of a parenthesis in a sequence of parentheses as the difference between the number of open parentheses and the number of closed parentheses in the prefix of the given sequence ending in that  parenthesis.
		\end{definition}
		The sequence of the depths of the parentheses coincides with the prefix sums of the sequence obtained by replacing every open parenthesis with a 1 and every closed parenthesis with a $-1$.
		
\begin{figure}[t]\centering 
			\includegraphics[width = 0.7\linewidth]{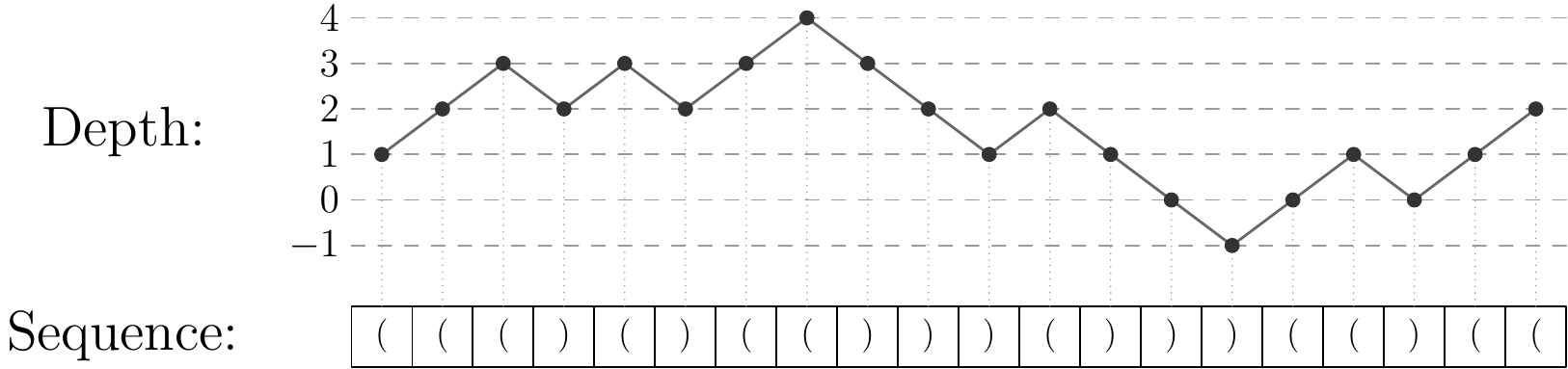}\\
			\caption{Depth of a sequence of parentheses. In this case, the summary of the sequence is the pair $(-1, 3)$. The summary of the first four parentheses is $(0, 2)$.\label{fig:basic summary}}

		\end{figure}

		\begin{definition}[summary of a sequence of parentheses]
		We define the \emph{summary} of a sequence of parentheses as the pair of integers $(a,b)$, where $a$ is the minimum between 0 and the minimum depth of the parentheses of the sequence, and $b$ is equal to the difference between the depth of the last parenthesis and $a$.
		\end{definition}
In the following, we  refer to the first value of the summary as to the \emph{down}-value, and to the second as to the \emph{up}-value. Note that the down-value of a summary is always non-positive, while the up-value is always non-negative. In Figure~\ref{fig:basic summary} we show a graphical representation of the depth of the parentheses in the sequence: for example, the summary of the whole sequence is the pair $(-1, 3)$, whilst the summary of the first four parentheses is $(0, 2)$.
		It should be clear that the summary of the sequence made of just one open parenthesis is $(0,1)$, while the summary of the sequence made of just one closed parenthesis is $(-1,0)$.
		
		The following lemmas hold for any sequence of parentheses:

\begin{restatable}{lemma}{monotonicity}
\label{lem:monotonicity}
	The down-values of the prefixes, taken in order, of any sequence of parentheses form a monotonically decreasing sequence of integers.
\end{restatable}
		
\begin{restatable}{lemma}{balancedseq}
			\label{lem:balancedseq}
			A sequence of parentheses is balanced if, and only if, its summary is equal to $(0,0)$. Any prefix of a balanced parenthetical sequence has down-value 0.
\end{restatable}

\begin{restatable}{lemma}{sumofsummaries}
\label{lem:sumofsummaries}
			Let $S_1, S_2$ be two sequences of parenthesis having summary $(a_1,b_1)$ and $(a_2,b_2)$ respectively. The summary of the sequence $S_1 + S_2$ obtained by concatenating $S_1$ and $S_2$ is the pair $(a_1,b_1) \boxplus (a_2,b_2)$, where the sum between summaries is defined as:
			$$\small (a_1, b_1) \boxplus (a_2, b_2) = \begin{cases}
				(a_1, b_1 + a_2 + b_2) & \text{if }b_1 + a_2 \ge 0\\
				(a_1 + b_1 + a_2, b_2) & \text{otherwise.}
			\end{cases}
			$$
\end{restatable}

\begin{restatable}{lemma}{summarysumassociativity}
			\label{lem:summarysumassociativity}
			The sum of two summaries defined above is an associative operation.
\end{restatable}
		
\begin{figure}[t]
\centering\includegraphics[width = .85\linewidth]{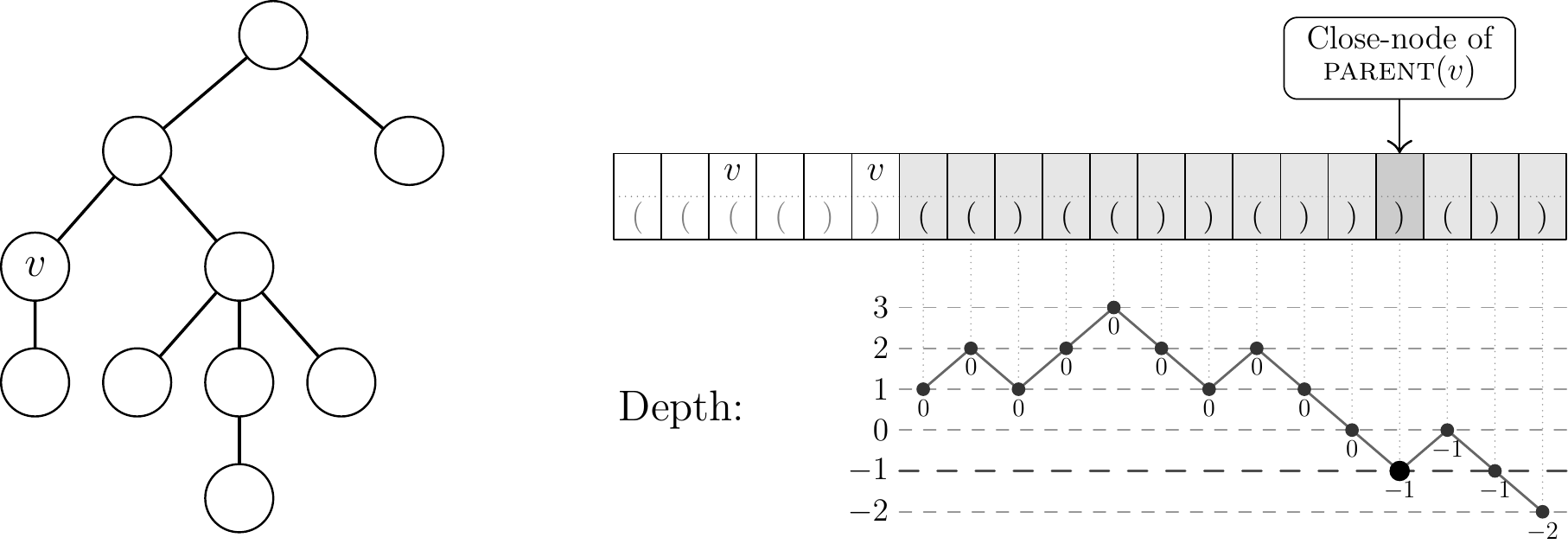}
\caption{Characterization of the parent of node $v$, as stated in Lemma~\ref{lem:lemfather}. The values under the small dots represent the down-values of the prefixes.\label{fig:father}}
\end{figure}

		As a consequence of Lemma~\ref{lem:summarysumassociativity},  as we mentioned before, we can store in each vertex of the BBST the sum of the summaries of all the vertices in its subtree. We proceed with the following lemma:
\begin{restatable}{lemma}{lemfather}
		\label{lem:lemfather}
		Let \emph{\var{close-v}} be the close-node associated with the non-root node $v$. The close-node associated with the parent of $v$ is the first (leftmost) node \emph{\var{u}} after \emph{\var{close-v}} reaching depth $-1$ relative to \emph{\var{close-v}}.
\end{restatable}
Lemma~\ref{lem:lemfather}, together with the associativity of $\boxplus$ and the monotonicity of the down values of the prefixes of any (sub)sequence of parentheses (Lemma~\ref{lem:monotonicity}), gives us an efficient way to locate the parent of any non-root node: we simply binary search the smallest prefix having a negative down-value, inside the suffix of the parenthetical sequence starting after \var{close-v}. Refer to figure~\ref{fig:father} for a visual insight. Similar properties hold for lca and ancestor: for example, for the $k$-th ancestor we can (binary) search the first node reaching relative depth $-k$ with respect to \var{close-v}, after \var{close-v}.
		
\section{Subtree (and path) operations}
\label{sec:operations}

In this section we detail the subtree and path operations. As we mentioned before, we assume that each node $v$ has an associated value (note that values can be generic objects, not only numbers), denoted by \textrm{val}$(v)$. We have the following three generic operations on a node that operate, respectively, on its children, on its subtree, and on the path from the node to the root:
		\begin{compactitem}
			\item \textsc{reduce-children}$(v, \oplus)$: Computes the value of 
			$$\textrm{val}(c_1) \oplus \cdots \oplus \textrm{val}(c_d),$$
			where $c_1, \ldots, c_d$ are the children of node $v$, and $\oplus$ is an associative operation (not necessarily invertible).
			\item \textsc{reduce-child-subtrees}$(v, \oplus, \otimes)$: Computes the value of $$\Sigma(c_1)\otimes\Sigma(c_2)\otimes\Sigma(c_3)\otimes\cdots\otimes\Sigma(c_d)$$
			where $c_1, \ldots, c_d$ are the children of node $v$, $\oplus$ and $\otimes$ are associative operations (not necessarily invertible), and $\Sigma(x) = \textrm{val}(x_1) \oplus \cdots \oplus \textrm{val}(x_m)$ is some information about the subtree rooted at $x$ and containing nodes $x_1, \ldots, x_m$.
			\item \textsc{combine}$(v, \odot)$: Computes the value of
			$$\textrm{val}(v_1) \odot \cdots \odot \textrm{val}(v_m),$$
			where $v = v_1, v_2, \ldots, v_m$ are the nodes in the path from $v$ to the root of the tree, and $\odot$ is an associative and invertible operation.
		\end{compactitem}
Differently from all other arguments, the operations denoted with $\oplus$, $\otimes$ and $\odot$ used in the three operations above have to be known in advance, so that the \dft\ knows what partial evaluations it should memoize in the nodes.

\begin{figure}[t]
\centering\includegraphics[width =\linewidth]{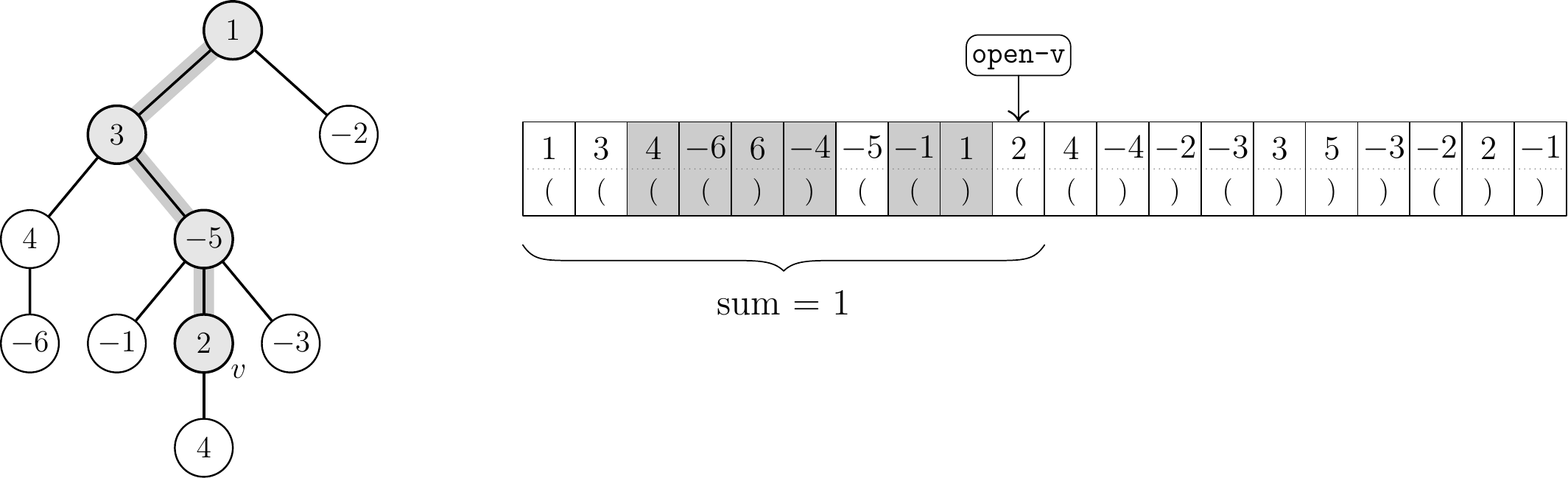}
\caption{Visual insight for Lemma~\ref{lem:combine}. The numbers written in the nodes of the tree on the left represent the values assigned to the vertices.\label{fig:combine}}
\end{figure}

Among the three operations, \textsc{combine} is the most straightforward, implementation-wise. The idea is to assign a value to both the open-nodes and close-nodes of the \dft: we assign the value of the vertex $\textrm{val}(v)$ to the open-node of $v$, and the opposite value $-\textrm{val}(v)$, i.e. the inverse of $\textrm{val}(v)$ with respect to operation $\odot$, to the corresponding close-node. We can thus state the following lemma, depicted in Figure~\ref{fig:combine} for the case $\odot$ is the traditional sum operator '$+$':

			\begin{restatable}{lemma}{combinelemma}
				\label{lem:combine}
Let \emph{\var{open-v}} be the open node associated with the tree node $v$. The value of $\textsc{combine}(v, \odot)$
				is equal to the $\odot$-combination of the values of the nodes in the prefix of the \dft ending in \emph{\var{open-v}}.
			\end{restatable}

In order to implement \textsc{reduce-children} and \textsc{reduce-child-subtree}, we need to extend the summary of a sequence of parentheses. 

Let us note that it is possible to uniquely decompose any sequence of parentheses in three contiguous (possibly empty) pieces, namely a \emph{prefix}, a \emph{body} and a \emph{suffix}. If the down-value of the sequence is (strictly) negative, then the prefix ends in leftmost minimal-depth parenthesis of the sequence, and the body ends in the rightmost minimal-depth parenthesis. If, on the contrary, the down-value of the sequence is 0, we can distinguish two separate cases: if the up-value is 0, then both the prefix and the suffix are empty, and the body coincides with the whole sequence; else, both the prefix and the body are empty, and the suffix coincides with the whole sequence.
In any case, notice that the body of a sequence is a balanced subsequence, made of zero or more \emph{subtrees}. As an example, consider these five sequences:
\begin{compactitem}
\item \texttt{)()(()}: the prefix is \texttt{)}, the body is \texttt{()} and the suffix is \texttt{(()}
\item \texttt{)())}: the prefix is \texttt{)())}, both body and suffix are empty
\item \texttt{))(}: the prefix is \texttt{))}, the body is empty and the suffix is \texttt{(}
\item \texttt{(()}: both the prefix and the body are empty, and the suffix is \texttt{(()}
\item \texttt{(()())}: both the prefix and the suffix are empty, while the body is \texttt{(()())}
\end{compactitem}
 We use this property, i.e. the unique decomposition of a sequence of parentheses, in the two summaries, used respectively by \textsc{reduce-children} and \textsc{reduce-child-subtree} to incrementally aggregate information about subtrees. Below we report the simpler one, used in \textsc{reduce-children}:
			\begin{definition}[rc-summary]
				An \emph{rc-summary} of a sequence of parentheses is a tuple having these fields:
				\begin{compactitem}
					\item {\var{prefix-depth}}, the depth of the minimal-depth parenthesis
					\item {\var{body-combination}}, the $\oplus$-combination of the values of the nodes associated with the subtrees of the body of the sequence.
					\item {\var{suffix-depth}}, the difference between the depth of the last parenthesis and the depth of any minimal-depth parenthesis.
					\item {\var{suffix-info}}, the value associated with the first node of the suffix, if any.
				\end{compactitem}
			\end{definition}

The similar \emph{rcs-summary}, used in \textsc{reduce-child-subtree}, is reported in the Appendix. These two summaries, to be stored as usual in the nodes of the underlying BBST, and the three generic functions above can be used to implement several functions, and below we report few examples.
\\[4mm]
\noindent\textbf{\textsf{Functions implemented using}} \textsc{reduce-children}\textbf{.}\quad
We can use \textsc{reduce-children} to implement:
		\begin{compactitem}
			\item \textsc{children-sum}$(v)$: Finds the sum of the values of the children of node $v$. This is equivalent to $\textsc{reduce-children}(v, +)$.
		 	\item \textsc{children-max}$(v)$: Finds the maximal value among those of the children of node $v$. This is equivalent to $\textsc{reduce-children}(v, \max)$.
		\end{compactitem}
Note that, if we set $\textrm{val}(x) = 1$ for every vertex in the forest,  $\textsc{degree}(v)$ can be derived as well from $\textsc{reduce-children}(v, +)$.
\\[4mm]
\noindent\textbf{\textsf{Functions implemented using}} \textsc{reduce-child-subtrees}\textbf{.}\quad
In the case of \textsc{reduce-child-subtrees} we can implement:
		\begin{compactitem}
			\item \textsc{subtree-sum}$(v)$: Finds the sum of the values of the nodes in the subtree of node $v$, and is equivalent to $\textrm{val}(v) + \textsc{reduce-child-subtrees}(v,+,+)$.
			\item \textsc{subtree-size}$(v)$: Finds how many nodes are there in the subtree of node $v$, and is equivalent to \textsc{subtree-sum}$(v)$ when $\textrm{val}(x) = 1$ for every node $x$ of the forest.
		 	\item \textsc{subtree-max}$(v)$: Finds the maximal value among those of the nodes in the subtree of node $v$, and is equivalent to $\max(\textrm{val}(v),\textsc{reduce-child-subtrees}(v, \max, \max)$.
			\item \textsc{maxsum-child}$(v)$: Finds the maximal value of \textsc{subtree-sum} among the children of node $v$. This is equivalent to $\textsc{reduce-child-subtrees}(v, +, \max))$. 	
		\end{compactitem}
\vspace{4mm}
\noindent\textbf{\textsf{Functions implemented using}} \textsc{combine}\textbf{.}\quad 		
		A simple example of \textsc{combine} is \textsc{depth}$(v)$, which finds the depth of node $v$, i.e. the distance from $v$ to the root of the tree $v$ belongs to. Indeed, this is equivalent to \textsc{combine}$(v, +)$, assuming $\textrm{val}(x) = 1$ for every node $x$ of the forest. We can implement \textsc{distance}$(u,v)$, i.e. the distance in the tree between $u$ and $v$, by computing $\text{\textsc{depth}}(u)+\textsc{depth}(v)-2\cdot\textsc{depth}(\textsc{lca}(u,v))$.

If we want to compute the distances in a weighted tree (i.e., we have weights on the edges), the same idea holds; since we store the information in the nodes, we store the weight of an edge connecting a child node to the parent node inside the child node. 

\section{Applications}
\label{sec:applications}
In this section we show, in order to provide a few examples, how to use \dfts\ to solve several problems that can be modeled as subtree problems. In particular, in all the applications that we describe we will refer to a common scenario: we ask queries about a single node $v$, and the queries can be answered by looking at the subtrees of $v$, i.e. the subtrees rooted in the children of $v$, together with the part of the tree that is above $v$, that we will denote by $\overline{T_v}$: this is the part of the tree that we reach through the parent of $v$. We will describe the applications in increasing order of complexity, from the perspective of the \dfts: indeed, as we will see, to compute the \emph{impact} of an articulation point $v$ we need to compute the size of the subtrees of $v$, and of $\overline{T_v}$; for the \emph{betweenness centrality} we also need to evaluate the sum of the squared sizes of the subtrees of $v$, and, finally, for the \emph{closeness centrality} we need the the sum of all the distances from $v$ to every node, both in its subtree and above it. % The    \emph{betweenness centrality} and \emph{closeness centrality} are detailed in Appendix~\ref{app:cc}.

\subsection{Biconnectivity properties and impact of articulation points}

\begin{figure}[t]
	\centering \raisebox{0.6cm}{ \includegraphics[scale=.54]{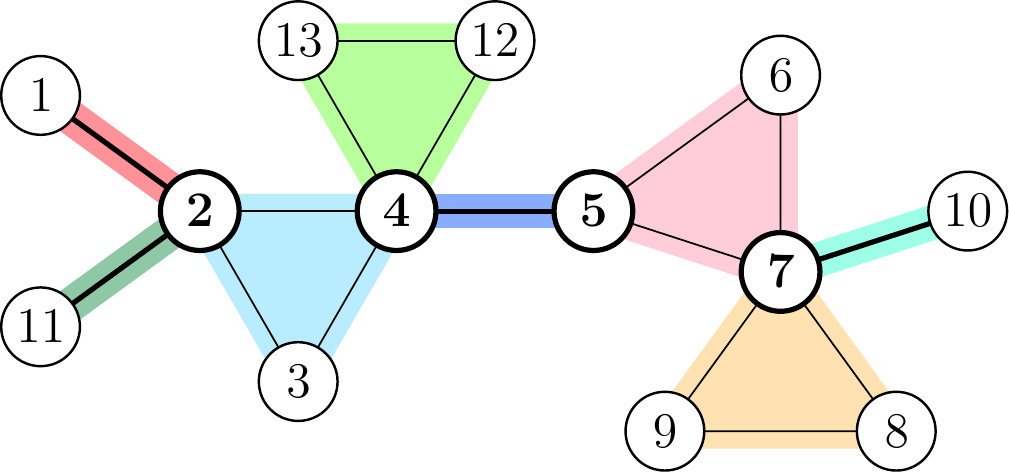}}\hfill 	\includegraphics[scale=.60]{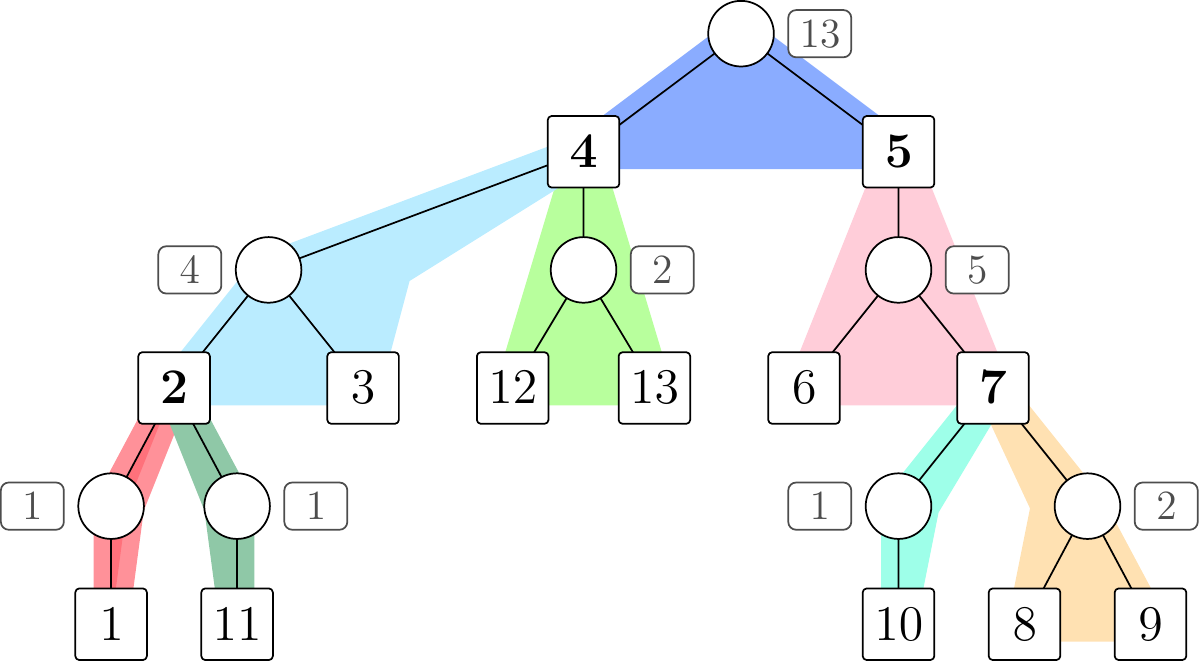}
			\caption{A graph (left) and its Block Forest~\cite{WT92} (right).\label{fig:BF}}
\end{figure}

The \dft\ can be used to maintain all the (bi)connectivity properties of a streaming graph, following the same approach proposed by Westbrook and Tarjan~\cite{WT92}: as we mentioned before, it is sufficient to observe that the \dft\ supports all the operations needed by the algorithm of Westbrook and Tarjan~ to maintain the Block Forest (shown in Figure~\ref{fig:BF}), including \textsc{condense} that, as we mentioned before,  is not a standard operation in the case of the \emph{dynamic tree} problem. Indeed, it is possible to maintain connected and biconnected components, and bridges and articulation points of a streaming graph. 

We now show how to answer queries on the \emph{impact} of an articulation point. We recall, from~\cite{AFL12trac}, that the impact of an articulation point $v$ is the number of nodes that get disconnected from the main connected component when $v$ is removed from the graph. Looking at the the Block Forest, Figure~\ref{fig:BF} (right), it is easy to see that the articulation points are exactly the square nodes that connect two or more round nodes (the biconnected components). When an articulation point is removed, its Block Tree splits into pieces: in order to compute the impact, we need to know the size of each of them: the impact is, by definition, the sum of all the size of the trees except the largest one (the main connected component). If we refer the subtree operations seen in the previous section, we can use the \dft\ in the following way:
\begin{compactitem}
\item The value in each round node in the tree is 0 (they corresponds to biconnected components), and 1 in each square node (corresponding to real nodes in the graph).
\item The size of the Block Tree can be computed by finding the root of the tree, using \textsc{root} and then computing its \textsc{subtree-size}.
\item The size of the maximum subtree of $v$ can be computed using \textsc{maxsum-child}.
\end{compactitem}
It is easy to see that, with the operations described above, we can compute the impact of a node, and thus we can state the following result.

\begin{restatable}{lemma}{lemimpact}
Using a \dft, it is possible to answer \emph{impact} queries of a vertex in time $\bigoh (\log n)$.
\end{restatable}

\subsection{Betweenness centrality}

The betweenness centrality definition involves shortest paths, but, since in  a tree there is exactly one path between each pair of nodes, the goal here is, given a vertex $v$, to count all the paths that pass through it. We can do this using \dfts\ in the following way. Let us assume that vertex $v$ has $k$ children, each of them with a corresponding subtree (eventually made by one node only, i.e. the child is a leaf). Let us denote with $st_1, st_2 \ldots st_k$ the subtrees of $v$. The number of (shortest) paths through $v$ can be partitioned into two  components: i) the paths between the subtrees of $v$ and the rest of the tree, i.e. $\{v\} \cup \overline{T_v}$, and ii) the paths between all the possible pairs of subtrees of $v$. The first component can be computed easily, using the fact that  $|\overline{T_v}| = \textsc{subtree-size}(\textsc{root}(v)) - \textsc{subtree-size}(v)$. The second component is the sum of the products of all the possible pairs of sizes, i.e., $\sum_{i \neq j} |st_i|\cdot|st_j|$; its computation is more tricky, if we want to avoid the iteration for every subtree. The idea is the following: 
\begin{compactitem}
				\item The value of each node in the tree is the pair $(1,1)$.
				\item We define $(a,a^2) \oplus (b, b^2)$ to be $(a+b, (a+b)^2)$.
				\item We have, as an invariant, that the values computed by  $\oplus$ are a couple made by a number and its square, e.g., $(x, x^2)$. Note that this defines an associative operation.
				\item We define $(a,a^2) \otimes (b,b^2)$ to be $(a+b, a^2 + b^2)$ (i.e., the usual vector sum). 
			\end{compactitem}
Now, if we call \textsc{reduce-child-subtrees}$(v, \oplus, \otimes)$ we obtain, for $v$, the couple made by the sum of the sizes of its subtrees, and by the sum of the squares of the sizes of its subtrees: $(|st_1| + |st_2| + \ldots + |st_k|,|st_1|^2 + |st_2|^2 + \ldots + |st_k|^2)=(\sum |st_i|, \sum |st_i|^2)$. It is easy to see, using the rule of the square of a sum, that the needed second component can be obtained by the couple of values. This allow us to state the following Lemma.
\begin{restatable}{lemma}{bc}
Using a \dft, it is possible to answer \emph{betweenness centrality} queries of a vertex  in time $\bigoh (\log n)$.
\end{restatable}

\subsection{Closeness centrality}
\label{sub:cc}
The closeness centrality~\cite{Bav50} of a vertex is defined as the reciprocal of its \emph{farness}, the sum of the distances to all the other vertices. We now show how to maintain the farness of each vertex, using the \dfts. The main ingredients are:
\begin{compactitem}
\item We modify the \dfts\ in order to support the two following operations: \textsc{add-to-path}$(v,\delta)$ that adds $\delta$ to all the vertices in the path between $v$ and the root, and \textsc{add-to-subtree}$(v,\delta)$ that adds $\delta$ to all the vertices in the subtree of $v$. Note that we can implement both these operations in $\bigoh (\log n)$ per update and value query, without affecting the complexity of the structural operations.
\item each vertex stores two values, \textsc{up-dists} that is the sum of the distances to the vertices in $\overline{T_v}$, and \textsc{down-dists} that is the sum of the distances to the vertices in its subtree. Note that the farness of a vertex is the sum of this two values. 
\end{compactitem}
Now, just to provide an example: assume that we are doing a \textsc{link} operation, adding the edge between $u$ and $v$, whose weight is $w$. Let us denote the size of the tree $u$ (resp. $v$) belongs to with $s_u$ (resp. $s_v$). The following operations need to be executed before the actual linking to maintain the information:
\begin{compactitem}
\item the \textsc{down-dists} of all the nodes in the path of $u$ are increased by $w\cdot\textsc{subtree-size}(v) + \textsc{down-dists}(v)$;
\item the \textsc{up-dists} of all the nodes in the subtree of $v$ (included) are increased by $w\cdot\textsc{subtree-size}(\textsc{root}(u)) + \textsc{up-dists}(u) + \textsc{down-dists}(v)$;
\item the \textsc{up-dists} of all the nodes in the tree containing $u$, with the only exception of the nodes in the path of $u$, are increased by $w\cdot\textsc{subtree-size}(v) + \textsc{down-dists}(v)$. In order to do so, we add it to all the nodes (i.e. the subtree of \textsc{root}$(u)$), and then we subtract it from all the nodes in the path of $u$. 
\end{compactitem}

The other structural update operations are similar, and can be derived in a similar fashion (we report them in the Appendix). This allow us to state the following Lemma.
\begin{restatable}{lemma}{bc}
Using a \dft, it is possible to answer \emph{closeness centrality} queries of a vertex in time $\bigoh (\log n)$.
\end{restatable}

\section{Conclusion and future works}
\label{sec:conclusion}
% Easy to understand, analyze and implement
% Oriented to subtree-operations (ST trees are more path-oriented)
% Extremely flexible
% We give a unified framework
% We give applications

In this paper we presented a novel data structure, the \emph{Depth First Tour Tree}. This structure is based on a linearization of a DFS visit of the tree, similarly to the ET-Trees (based on a Euler Tour). 

The structure is simple and easy to implement; it provides a framework for a large class of data aggregation tasks -- especially on subtrees, a task that is usually non-trivial with other data structures. 
Furthermore, \dfts\ can naturally aggregate on all the children subtrees of a node $v$ in parallel without having to pay a cost proportional to the degree of $v$ itself: as we already mentioned, given a node $v$ it takes $\bigoh(\log n)$, independently from the degree of $v$, to answer the child of $v$ whose subtree is the largest.

This flexibility, related to subtree queries, is paid by the \textsc{evert} operation, that requires a cost proportional to the depth of the node being everted. However, as discussed, when either the number of eversions is small compared to the total number of queries performed, or the costs of the eversion is amortized, the cost of \textsc{evert} can be regarded as being $\bigoh(\log n)$ like all the other structural operations. 

We showed that this is the case in all the applications presented in the previous section. We described how to customize the data structure in order to answer queries for three different applications: impact of the removal of an articulation point from a graph, betweenness centrality and closeness centrality of a dynamic tree. 

In the future, we plan to experimentally assess the performance of our data structure, and compare it with the existing alternatives, following the approach of \cite{TW09}. We believe that the simplicity of our approach, when compared e.g. to the work of \cite{Radzik98implementationof} in the context of the \textsc{subtree-max} operation, is likely to deliver faster and more readable code in practice.

\vfill 
\pagebreak

%----------------------------------------------------------------------------------------
%	REFERENCE LIST
%----------------------------------------------------------------------------------------

\bibliography{dynamic-tree}

\vfill 
\pagebreak
%----------------------------------------------------------------------------------------
\appendix

\section{Implementation of \dfts\ using Splay Trees}
\label{app:implementation}

In this appendix we detail the pseudo-code for all the supported operations in a \dft, using the Splay Trees~\cite{ST85}, that are used by Tarjan and  and Tarjan~\cite{TW09} to implement both the ST-trees~\cite{ST83,ST85}, ET-trees~\cite{HK99,Tar97}.

The \dft\ is thus stored as an augmented splay tree, where the comparison $x \prec y$ between two entries $x$ and $y$ of the depth first tour evaluates to \textsc{true} iff entry $x$ comes before entry $y$ in normal left-to-right order.

Since the focus of the paper has been devoted to subtree computations, we note here that in \Cref{sub:combine} we show an example of how to use \textsc{combine} to compute a path operation.

\subsection{Basic splay operations}

		We will take for granted the implementation of these basic operations on the splay tree, besides the  tree rotations, \textsc{splay}, \textsc{splay-erase}, \textsc{splay-min} / \textsc{splay-max} and \textsc{splay-predecessor} / \textsc{splay-successor}:
		\begin{description}
			\item[\textsc{splay-root}$(v)$:] Returns the root node of the splay tree containing node $v$.
			
			\item[\textsc{splay-lca}$(u,v)$:] Returns the lowest common ancestor of the splay nodes $u$ and $v$. Of course, $u$ and $v$ must belong to the same splay tree (i.e. \textsc{splay-root}$(u)$ = \textsc{splay-root}$(v)$).
			
			\item[\textsc{splay-merge}$(u, v)$:] Joins the splay tree $T_1$ containing node $u$ with the splay tree $T_2$ containing node $v$. If $u$ and $v$ belong to the same tree, nothing happens. If $u$ and $v$ belong to different tree, the keys contained in $T_1$ are considered to precede all the keys in $T_2$.
			
			\item[\textsc{splay-split}$(v)$:] Splits the splay tree $T$ containing $v$ into two different splay trees: the first contains all the keys which are $\preceq v$, and the second contains all the keys which are $\succ v$.
			
			\item[\textsc{splay-precedes}$(u,v)$]: Returns \textsc{true} if $u \preceq v$, \textsc{false} otherwise.
		\end{description}
		Operation \textsc{splay-root} can be implemented by simply moving from a node to its parent until we eventually reach the root of the splay tree. This method clearly results in amortized logarithmic complexity with respect to the tree size.
		
		\textsc{splay-lca} can be implemented by marking all the nodes in the path from $v$ to the root, and then moving up the tree starting from $u$, stopping at the first marked node found on this path, which corresponds to the sought ancestor.
		
Also, it is possible   (see \cite{GKSTW11})  to support \textsc{splay-merge} and \textsc{splay-split} in logarithmic time in the size of the trees involved.
		
		Implementation for \textsc{precedes} is given in \Cref{algo:precedes}.
		\begin{algorithm}[H]
		  \small
		  \caption{\small Implementation of \textsc{splay-precedes}}
		  \label{algo:precedes}
		  \begin{algorithmic}[1]
		    \Procedure{splay-precedes}{$u, v$}\Comment{$u$ and $v$ are dft nodes.}
		    \State \var{successor} $\gets$ \textsc{splay-successor}$(u)$
			\State \textsc{splay-split}$(u)$
			\State \var{answer} $\gets$ (\textsc{spay-root}($u$) == \textsc{splay-root}($v$))
			\If{\var{successor} $\neq$ \textsc{null}}\Comment{Restore tree.}
				\State \textsc{splay-merge}($u$, \var{successor})
			\EndIf
			\State \textbf{return} \var{answer}
		    \EndProcedure
		  \end{algorithmic}
		\end{algorithm}
		
		We will assume that every splay node contains a pointer to its \emph{twin}, i.e. to the other dft node associated to the same tree node.
		
In general, we will maintain a collection of disjoint splay trees, where in turn a splay tree can maintain the depth first tours of one or more (disjoint) trees. When a splay tree contains only one dft, we say that the dft has a \emph{dedicated} splay tree. We provide an internal operation, \textsc{splice}$(v)$, which makes sure that the dft of the tree containing $v$ gets a dedicated splay tree. Notice that \textsc{splice} alters the internal splay tree representation, without affecting the represented tree. Assuming that we already have implemented operation \textsc{root}, implementing \textsc{splice} in logarithmic time is rather straightforward and is done in \Cref{algo:splice}.
		\begin{algorithm}[H]
		  \small
		  \caption{\small Implementation of \textsc{splice}}
		  \label{algo:splice}
		  \begin{algorithmic}[1]
		    \Procedure{splice}{$v$}
		    \State \var{open-root} $\gets$ open-node of \textsc{root}$(v)$
		    \State \var{close-root} $\gets$ close-node of \textsc{root}$(v)$
		    \State \var{predecessor} $\gets$ \textsc{splay-predecessor}$(\var{open-root})$
		    \If{\var{predecessor} $\neq$ \textsc{null}}
		    	\State \textsc{splay-split}(\var{predecessor})
		    \EndIf
			\State \textsc{splay-split}(\var{close-root})
		    \EndProcedure
		  \end{algorithmic}
		\end{algorithm}

\subsection{Import/export operations}

Building the \dft\ of a given tree, encoded in the adjacency list format, is a very simple task, and can be seen as an easy modification of the classical dfs algorithm.
		
		The opposite task, i.e., restoring the original tree given its depth first tour, is also very simple. Indeed, it is enough to keep track of the current open node using a stack, while we process every node in the given \dft: see~\Cref{alg:conversion}.
		\begin{algorithm}[H]
		  \small
		  \caption{\small Depth first tour to tree conversion\label{alg:conversion}}
		  \begin{algorithmic}[1]
		    \Procedure{dft-to-tree}{\var{DFT}}\Comment{\var{DFT} is a list here}
		    \State \var{s} $\gets$ empty stack
		    \ForAll{(\var{node}, \var{tag}) \textbf{in} \var{DFT} \textbf{in order},}
		      \If{\var{tag} is an \emph{open-tag}}
			    \If{\var{s}.empty()}
	              \State mark \var{node} as the root of the tree
			    \Else
			      \State add \var{node} to the children of \var{s}.top().
			    \EndIf
			    \State \var{s}.push(\var{node})
		      \Else
		        \State \var{s}.pop()
		      \EndIf
		    \EndFor
		    \EndProcedure
		  \end{algorithmic}
		\end{algorithm}

		To perform \textsc{import-tree} we first construct the depth first tour of the input tree, and then build a splay tree corresponding to it. Since the order of the nodes in the depth first tour coincides with the order maintained by the underlying splay tree, we can perform a linear time tree construction as described in [...]. To correctly maintain the extra information stored in the nodes of the splay tree, we can propagate them from the leaves up to the root, combining them using the \textsc{recalc-extra-info} function, leading to an overhead which is linear in the size of tree, hence not affecting the total complexity of the operation.
		
		Operation \textsc{export-tree} performs an in-order traversal of the (spliced) splay tree, extracting a list version of the depth first tour it represents, and then runs \textsc{dft-to-tree} on it. Since both operations have linear complexity in the tree size, we can support \textsc{export-tree} in linear time.
		
		\subsection{Structural updates}
		
In this section we describe the implementation of the structural update operations on a \dft. In particular, the most important operations are the \textsc{link} and \textsc{cut}, whose effect on the parenthetical sequence is shown in \Cref{fig:linkcut}.

Suppose an edge is created between the root $v$ of tree $T_2$ and node $u$ of tree $T_1$. From the point of view of depth first tours, what happens is that the dft of $T_2$ is inserted into the dft of $T_1$ right after the open-node corresponding to $u$. See \Cref{alg:link} below.

		\begin{algorithm}[H]
		  \small
		  \caption{\small Implementation of \textsc{link} \label{alg:link}}
		  \begin{algorithmic}[1]
		    \Procedure{link}{$u,v$}
		    \If{\textbf{not} \textsc{same-tree}($u, v$)}
			    \State \var{open-u} $\gets$ open-node of node $u$ in the dft
			   	\State \var{close-u} $\gets$ close-node of node $u$ in the dft
			   	\State \var{open-v} $\gets$ open-node of node $v$ in the dft
		    	\State \textsc{splay-split}(\var{open-u})
		    	\State \textsc{splay-merge}(\var{open-u}, \var{open-v})
		    	\State \textsc{splay-merge}(\var{open-u}, \var{close-u})
		    \EndIf
		    \EndProcedure
		  \end{algorithmic}
		\end{algorithm}	
		
		Operation \textsc{cut}$(v)$ is analogous and has the effect of extracting the sub-segment of the dft corresponding to the subtree rooted in $v$, as illustrated in \Cref{fig:linkcut}. Its implementation is symmetric to the one of \textsc{link}:
		\begin{algorithm}[H]
		  \small
		  \caption{\small Implementation of \textsc{cut}}
		  \begin{algorithmic}[1]
		    \Procedure{cut}{$v$}
		    \State \var{root} $\gets$ \textsc{root}($v$)
		    \If{$v$ $\neq$ \var{root}}
				\State \var{open-v} $\gets$ open-node of node $v$ in the dft
				\State \var{close-v} $\gets$ close-node of node $v$ in the dft
				\State \var{open-root} $\gets$ open-node of \var{root} in the dft
				\State \var{close-root} $\gets$ close-node of \var{root} in the dft

				\State \textsc{splay-split}(\textsc{splay-predecessor}(\var{open-v}))
				\State \textsc{splay-split}(\var{close-v})
				\State \textsc{splay-merge}(\var{open-root}, \var{close-root})
		    \EndIf
		    \EndProcedure
		  \end{algorithmic}
		\end{algorithm}	
		Note that the call to \textsc{predecessor} in line 8 is licit: since $v$ is not the root of the tree, \var{open-v} cannot be the first node in the dft.
		
\begin{figure}[t]\centering 
	\includegraphics[width = \linewidth]{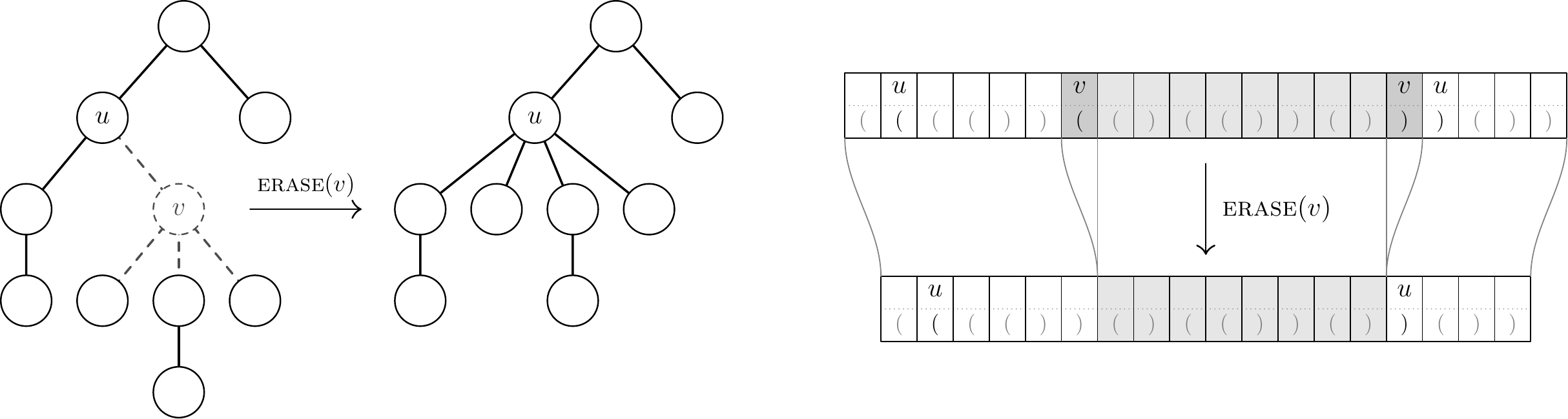}
			\caption{Effects of the \textsc{condense} operation on the dft.\label{fig:condense}}
		\end{figure}

		The effect of operation \textsc{condense}($v$) on the dft of the tree is explored in \Cref{fig:condense}, and corresponds to the deletion of the open- and close-node associated with $v$ in the dft.
		\begin{algorithm}[H]
		  \small
		  \caption{\small Implementation of \textsc{condense}}
		  \begin{algorithmic}[1]
		    \Procedure{condense}{$v$}
			\State \var{open-v} $\gets$ open-node of node $v$ in the dft
			\State \var{close-v} $\gets$ close-node of node $v$ in the dft
			\State \textsc{splay-erase}(\var{open-v})
			\State \textsc{splay-erase}(\var{close-v})
		    \EndProcedure
		  \end{algorithmic}
		\end{algorithm}

		Operation $\textsc{erase}(v)$ is equivalent to a call to $\textsc{cut}(v)$ followed by a call $\textsc{condense}(v)$.
		\begin{algorithm}[H]
		  \small
		  \caption{\small Implementation of \textsc{erase}}
		  \begin{algorithmic}[1]
		    \Procedure{erase}{$v$}
			\State \textsc{cut}($v$)
			\State \textsc{condense}($v$)
		    \EndProcedure
		  \end{algorithmic}
		\end{algorithm}		 
		
Notice that both \textsc{erase} and \textsc{condense} may lead to dft having non-dedicated splay trees.
		
		Operation \textsc{evert}$(v)$ can be implemented in two different ways. The first one makes a call to \textsc{export-tree}, operates an $\bigoh(n)$ evert operation on the adjacency list version of the tree and finally rebuilds the splay version using \textsc{import-tree}, for a total of $\bigoh(n)$ operations on a tree of size $n$. The second way of performing the eversion consists in the following recursive algorithm, whose complexity is $\bigoh(h \log n)$, where $h$ is the depth of node $v$:
		\begin{algorithm}[H]
		  \small
		  \caption{\small Implementation of \textsc{evert}}
		  \begin{algorithmic}[1]
		    \Procedure{evert}{$v$}
		    \State \var{root} $\gets$ \textsc{root}($v$)
			\If{$v \neq \var{root}$}
				\State \var{parent} $\gets$ \textsc{parent}($v$)
				\State \textsc{cut}($v$)
				\State \textsc{evert}(\var{parent})
				\State \textsc{link}($v$, \var{parent})
			\EndIf
		    \EndProcedure
		  \end{algorithmic}
		\end{algorithm}		
		
		\subsection{Non-structural operations}
		Operation \textsc{same-tree}$(u,v)$ is straightforward and corresponds to checking whether $\textsc{root}(u) = \textsc{root}(v)$ or not.
		
		To implement \textsc{is-descendant} we first make the following observation:
		\begin{lemma}
			\item Let $u$ and $v$ be two nodes, having open-nodes {\var{open-u}}, \emph{\var{open-v}} and close-nodes \emph{\var{close-u}}, \emph{\var{close-v}} respectively. Node $u$ is a descendant of node $v$ if and only if $\emph{\var{open-v}} \preceq \emph{\var{open-u}}$ and $\emph{\var{close-u}} \preceq \emph{\var{close-v}}$.
		\end{lemma}
		Using the previous observation, implementing \textsc{is-descendant} becomes a straightforward task, shown in \Cref{algo:is descendant}.
		\begin{algorithm}[H]
		  \small
		  \caption{\small Implementation of \textsc{is-descendant}}
		  \label{algo:is descendant}
		  \begin{algorithmic}[1]
		    \Procedure{is-descendant}{$u, v$}
			    \State \var{open-u} $\gets$ open-node of node $u$ in the dft
			    \State \var{close-u} $\gets$ close-node of node $u$ in the dft
			    \State \var{open-v} $\gets$ open-node of node $v$ in the dft
			    \State \var{close-v} $\gets$ close-node of node $v$ in the dft
			    \State \textbf{return} \textsc{splay-precedes}(\var{open-v}, \var{open-u}) $\land$ \textsc{splay-precedes}(\var{close-u}, \var{close-v})
		    \EndProcedure
		  \end{algorithmic}
		\end{algorithm}
		
		Operation \textsc{list-children} repeatedly uses operation \textsc{splay-successor} to traverse consecutive siblings, shown in \Cref{algo:list children}.
		
		\begin{algorithm}[H]
		  \small
		  \caption{\small Implementation of \textsc{list-children}}
		  \label{algo:list children}
		  \begin{algorithmic}[1]
		    \Procedure{list-children}{$v$}
			    \State \var{open-v} $\gets$ open-node of node $v$ in the dft
				\State \var{close-v} $\gets$ close-node of node $v$ in the dft
			    \State \var{current} $\gets$ \textsc{splay-successor}(\var{open-v})
			    \State \var{children} $\gets$ empty list
			    \While{\var{current} $\neq$ \var{close-v}}
			    	\State \var{children}.push(tree node associated to \var{current})
			    	\State \var{current} $\gets$ \textsc{splay-successor}(\var{current.twin})
			    \EndWhile
			    \State \textbf{return} \var{children}
		    \EndProcedure
		  \end{algorithmic}
		  Note: we recall that the twin of a dft node $u$ is the (pointer to) the other dft node $u'$ associated to the same tree node as $u$. In this case, line 8 finds the next sibling of the tree node associate with \var{current}.
		\end{algorithm}	
		\vspace{2mm}
		
		Operation \textsc{parent}, briefly described in \Cref{sec:dft} is the first non-trivial operation, as we begin to exploit the parenthetical sequence of dft and to work on the augmented splay tree nodes. As such, we first need to set some definitions about sequences of parentheses. %\nota{chiarire che si può fare in O(1) con spazio addizionale se si rinuncia al condense\gabriele{Citare i condensible nodes di Tarjan}}
		\begin{lemma}
			The suffix of the dft of the whole tree starting after \var{close-v} begins with the concatenation of the dft of zero or more siblings of node $v$, followed by the close-node of the parent of $v$. 
		\end{lemma}
		
\begin{figure}[t]

\centering
			\includegraphics[width = .9\linewidth]{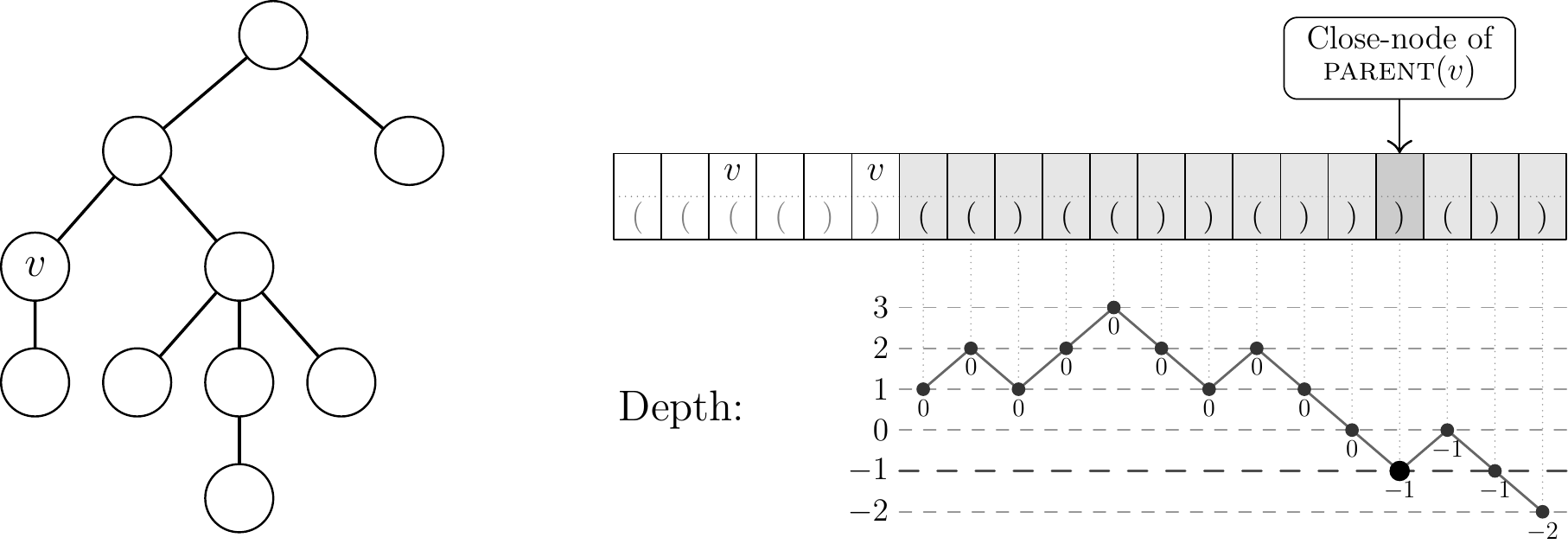}\\
			\caption{Characterization of the close-node of the parent of $v$. The small numbers under the dots represent the summary down-values.\label{fig:close father}}
		\end{figure}

		We provide a visual insight in \Cref{fig:close father}.
		
		Given the monotonicity of the summary down-values noted above, we can devise a binary search algorithm for finding the parent of any non-root node, shown in \Cref{algo:father}.

			\begin{algorithm}[H]
			  \small
			  \caption{\small Implementation of \textsc{parent}}
			  \label{algo:father}
			  \begin{algorithmic}[1]
			  	\Procedure{recursive-parent}{\var{splay-node}, \var{summary}}\Comment{pre: the down-value of summary is 0}
			  		\If{\var{splay-node.left-child} $\neq$ \textsc{null}}
			  			\If {(\var{summary} $\boxplus$ \var{splay-node.left-child.range-summary}).\var{down-value} $\le -1$}
			  				\State \textbf{return} \textsc{recursive-parent}(\var{splay-node.left-child}, \var{summary})
			  			\Else
			  				\State \var{summary} $\gets$ \var{summary} $\boxplus$ \var{splay-node.left-child.range-summary}
			  			\EndIf
			  		\EndIf
			  		\If{(\var{summary} $\boxplus$ \var{splay-node.node-summary}).\var{down-value} $\le -1$}
			  			\State \textbf{return} \var{splay-node}
			  		\Else
			  			\State \var{summary} $\gets$ \var{summary} $\boxplus$ \var{splay-node.node-summary}
			  		\EndIf
			  		\State \textbf{return} \textsc{recursive-parent}(\var{splay-node.right-child}, \var{summary})
			  	\EndProcedure
				\State
			    \Procedure{parent}{$v$}
			    \State \var{root} $\gets$ \textsc{root}($v$)
				\If{$v \neq \var{root}$}
					\State \var{close-v} $\gets$ close-node of $v$
					\State \var{successor} $\gets$ \textsc{splay-successor}(\var{close-v})
					\State \textsc{splay-split}(\var{close-v})
					\State \var{close-parent} $\gets$ \textsc{recursive-parent}(\textsc{splay-root}(\var{successor}), $(0,0)$)
					\State \textsc{splay-merge}(\var{close-v}, \var{successor})
					\State \textbf{return} the tree node having \var{close-parent} as close-node
				\Else
					\State \textbf{return} \textsc{null}
				\EndIf
			    \EndProcedure
			  \end{algorithmic}
			\end{algorithm}

			Operation \textsc{lca} can be supported in a similar fashion, since the following result holds:
			\begin{lemma}[characterization of the lca]
				\label{prop:characterization of lca}
				Let $u,v$ be distinct nodes belonging to the same tree, for which none is a descendant of the other, and let \emph{\var{close-u}} and \emph{\var{close-v}} be their close-nodes in the dft. Suppose further, without loss of generality, that $\emph{\var{close-u}} \prec \emph{\var{close-v}}$. Consider the subsequence of the parenthetical sequence of $T$, starting in \emph{\var{close-u}} and ending in \emph{\var{close-v}}, and let $w$ be the leftmost dft-node having minimal depth. The lowest common ancestor of $u$ and $v$ is the parent $a$ of the tree node corresponding to $w$. More specifically, $w$ is child of $a$ closest to node $u$, i.e. the second-to-last node in the path from $u$ to $a$.
			\end{lemma}
			
\begin{figure}[t]

				\includegraphics[width = \linewidth]{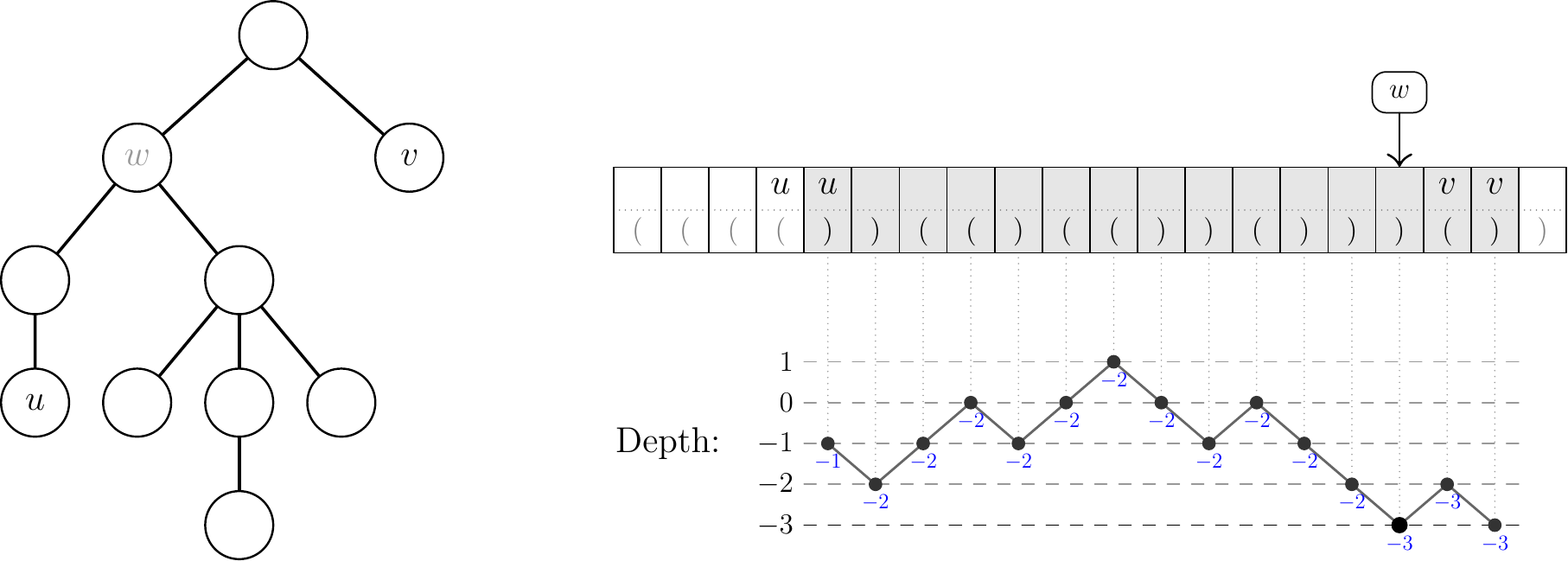}\\
				\caption{Characterization of the lca of nodes $u$ and $w$. The lca is the parent of $w$, the node associated with any parenthesis having minimal depth.\label{fig:lca}}

			\end{figure}

			See \Cref{fig:lca} for a visual insight. To quickly determine $w$ we augment the concept of summary, so that it keeps track of some parenthesis reaching minimal depth. More formally, we consider the following definition:
			\begin{definition}\hspace{0mm}\emph{\textbf{(lca-summary of a sequence of parenthesis)}}
				The \emph{lca-summary} of a sequence of parentheses is an ordered pair $(s,p)$, where $s$ is the summary of the given sequence and $p$ is a (pointer) reference to the leftmost parenthesis having depth equal to the down-value of $s$. If no such parenthesis exists, $p$ is set to \textsc{null}.
			\end{definition}
			
			It is easy to adapt the addition operator between summaries to lca-summaries, so that we can easily evaluate the lca-summary of the concatenation of two sequences, as can be seen in \Cref{prop:sum of lca-summaries}.

			\begin{lemma}
				\label{prop:sum of lca-summaries}
				Let $S_1, S_2$ be two sequences of parenthesis having lca-summary $(s_1=(a_1,b_1),p_1)$ and $(s_2=(a_2,b_2),p_2)$ respectively. The summary of the sequence $S_1 + S_2$ obtained by concatenating $S_1$ and $S_2$ is the pair $(s_1,p_1) \boxplus (s_2,p_2)$, where the sum between lca-summaries is defined as:
				$$ (s_1,p_1) \boxplus (s_2,p_2) = \begin{cases}
					(s_1 \boxplus s_2, p_1) & \text{if }b_1 + a_2 \ge 0\\
					(s_1 \boxplus s_1, p_2) & \text{otherwise.}
				\end{cases}
				$$
			\end{lemma}
			\begin{lemma}
				\label{prop:pointer to null}
				Let $(s, p)$ be the lca-summary of a sequence of parentheses. Pointer $p$ points to \textsc{null} if and only if all the depths of the parentheses are (strictly) positive.
			\end{lemma}
			\begin{lemma}
				\label{prop:lca-summary sum associativity}
				The sum of two lca-summaries defined above is an associative operation.
			\end{lemma}
			
			Note that by Proposition~\ref{prop:pointer to null} it follows that the lca-summary associated with the range indicated in \Cref{prop:characterization of lca} has a non-\textsc{null} reference, since the first parenthesis of the range is a closed-parenthesis.
			
			As before, we augment the nodes of the splay tree so that every node keeps the extra values
			\begin{compactitem}
				\item \var{node-lca-summary}, corresponding to the lca-summary of the node in question;
				\item \var{range-lca-summary}, the lca-summary of the subsequence associated with the splay subtree rooted in the node in question.
			\end{compactitem}
			
			We sketch the algorithm for determining the lca of two nodes in \Cref{algo:lca}.
			\begin{algorithm}[H]
			  \small
			  \caption{\small Implementation of \textsc{lca}}
			  \label{algo:lca}
			  \begin{algorithmic}[1]
			    \Procedure{lca}{$u,v$}
				    \If{\textsc{is-descendant}$(u,v)$}
				    	\State \textbf{return} $v$
				    \State\hspace{-\algorithmicindent}\textbf{elseif} \textsc{is-descendant}$(v,u)$ \textbf{then}
				    	\State \textbf{return} $u$
				    \EndIf
				    \State \var{close-u} $\gets$ close-node of node $u$ in the dft
				    \State \var{close-v} $\gets$ close-node of node $v$ in the dft
				    \If{\textsc{splay-precedes}(\var{close-v},\var{close-u})}
				    	\State \textbf{return} \textsc{lca}($v,u$)
				    \Else
				    	\State \var{pred-u} $\gets$ \textsc{splay-predecessor}(\var{close-u})
				    	\State \var{succ-v} $\gets$ \textsc{splay-successor}(\var{close-v})
				    	\State \textsc{splay-split}(\var{pred-u})
				    	\State \textsc{splay-split}(\var{close-v})
				   		\State \var{range-root} $\gets$ \textsc{splay-root}(\var{close-v})
				   		\State \var{dft-w} $\gets$ \var{range-root}.\var{range-lca-summary}.\var{p}
				   		\State \textsc{splay-merge}(\var{pred-u}, \var{close-u})
				   		\State \textsc{splay-merge}(\var{close-u}, \var{succ-v})
				   		\State $w$ $\gets$ the tree node associated with \var{dft-w}
				   		\State \textbf{return} \textsc{parent}($w$)
				    \EndIf
			    \EndProcedure
			  \end{algorithmic}
			  \textbf{Note:} note that, since lines 12-19 run only if $u$ is not a descendant of $v$ and $v$ is not a descendant of $u$, \var{prec-u} and \var{succ-v} are non-\textsc{null}, well-defined nodes.
			\end{algorithm}	
			
We conclude this section discussing how to implement \textsc{root}. One may be tempted to say that $\textsc{root}(v)$ is the node associated with the $\textsc{splay-min}$ of the splay tree containing the dft nodes corresponding to $v$. Unfortunately, this is not true when the dft of the tree containing $v$ is kept in a non-dedicated splay tree. Thus we need the following in \Cref{prop:characterization of root}.
			\begin{lemma}[characterization of the root]
				\label{prop:characterization of root}
				Let $v$ be a node, and let \emph{\var{close-v}} be the close-node associated with $v$. The close-node of the root of the tree containing $v$ is the leftmost dft-node $\succeq$ \emph{\var{close-v}} having minimal depth.
			\end{lemma}

In other words, \Cref{prop:characterization of root} states that the $p$ value of the lca-summary of the suffix of the splay tree starting in \var{close-v} is the close-node of the root of $v$. As before, \Cref{prop:pointer to null} guarantees that the $p$ value of that range is not \textsc{null}, as the first node in the range is a closed parenthesis. This leads to an easy implementation, shown in \Cref{algo:root}.

			\begin{algorithm}[H]
			  \small
			  \caption{\small Implementation of \textsc{root}}
			  \label{algo:root}
			  \begin{algorithmic}[1]
			    \Procedure{root}{$v$}
					\State \var{close-v} $\gets$ close-node of node $v$ in the dft
				    \State \var{predecessor} $\gets$ \textsc{splay-predecessor}(\var{close-v})
				    \State \textsc{splay-split}(\var{predecessor})
				    \State \var{splay-root} $\gets$ \textsc{splay-root}(\var{close-v})
				    \State \var{dft-w} $\gets$ \var{range-root.range-lca-summary.p}
				    \State \textsc{splay-merge}(\var{predecessor}, \var{close-v})
				    \State $w$ $\gets$ the tree node associated with \var{dft-w}
				    \State \textbf{return} $w$
			    \EndProcedure
			  \end{algorithmic}
			\end{algorithm}

		\subsection{Reductions and combinations}
\label{sub:combine}
			
			We recall that operation \textsc{combine}  value of $$\textrm{val}(v_1) \odot \textrm{val}(v_2) \odot \cdots \odot \textrm{val}(v_h),$$
			where $v = v_1, v_2,\ldots, v_h$ are the nodes in the path from $v$ to the root of the tree, and $\odot$ is any invertible associative binary operation acting on the values attached to the nodes. We augment the splay tree, adding two fields:
			\begin{compactitem}
				\item \var{item-val}, the value of the node, and
				\item \var{range-val}, the $\odot$-combined value of \var{item-val} for all the dft nodes in the splay subtree rooted in the node in question
			\end{compactitem}
			In particular, if \var{v} is a dft node associated with the tree node $v$, we set
			$$
			\var{v.item-val} = \begin{cases}
				\textrm{val}(v) & \text{ if \var{v} is a dft open-node}\\
				-\textrm{val}(v) & \text{ if \var{v} is a dft close-node}
			\end{cases}
			$$
			where $-x$ indicates the inverse of $x$ with respect to $\odot$.
			\begin{restatable}{lemma}{combinelemma2}
				\label{prop:combine}
Let \emph{\var{open-v}} be the open node associated with the tree node $v$. The value of $\textsc{combine}(\odot, v)$
				is equal to the $\odot$-combination of the \emph{\var{item-val}} of the nodes in the prefix of the dft ending in \emph{\var{open-v}}.
			\end{restatable}
			
\begin{figure}[t!]
				\includegraphics[width = \linewidth]{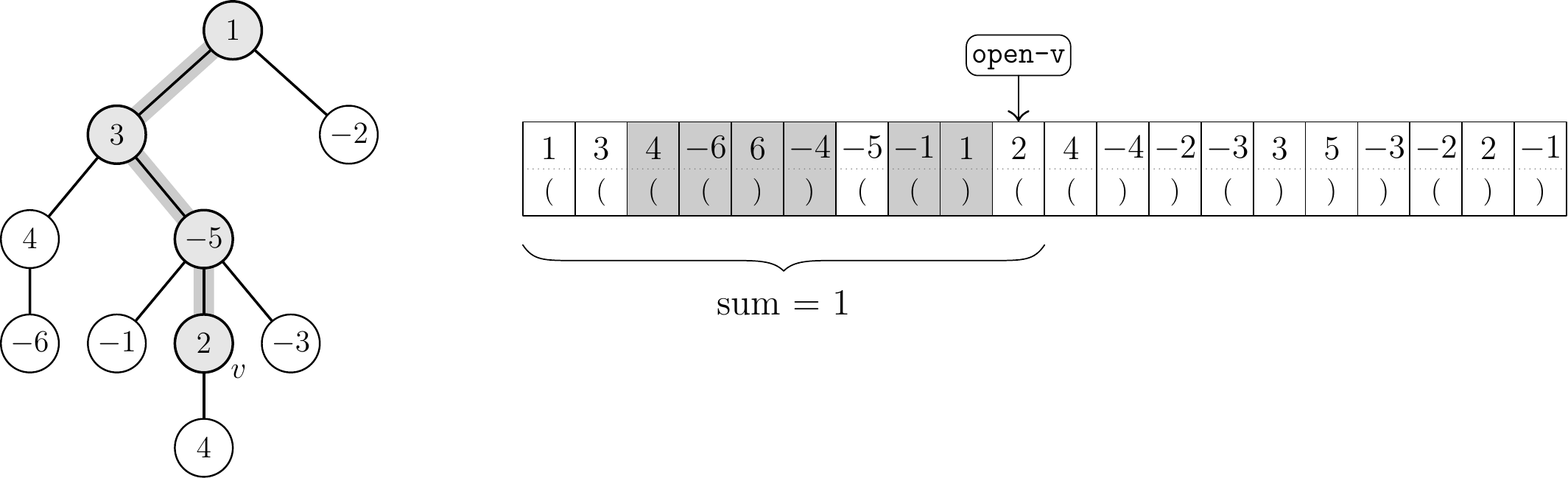}
				\caption{Visual insight for \Cref{prop:combine}\label{fig:combine example}. The numbers written in the nodes of the tree on the left represent the values assigned to the vertices.}
			\end{figure}

As an example, consider the case in which $\odot$ denotes the usual addition of real numbers: a visual insight for \Cref{prop:combine} is given in \Cref{fig:combine example}. The pseudocode of \textsc{combine} is detailed in \Cref{algo:combine}.

			\begin{algorithm}[H]
			  \small
			  \caption{\small Implementation of \textsc{combine}}
			  \label{algo:combine}
			  \begin{algorithmic}[1]
			    \Procedure{combine}{$\odot,v$}
				    \State \var{open-v} $\gets$ open-node of node $v$ in the dft
					\State \var{close-root} $\gets$ close-node of the tree root
					\State \textsc{splay-split}(\var{close-v})
					\State \var{answer} $\gets$ \textsc{splay-root}(\var{close-v}).\var{range-val}
					\State \textsc{splay-merge}(\var{close-v}, \var{close-root})
					\State \textbf{return} \var{answer}
			    \EndProcedure
			  \end{algorithmic}
			\end{algorithm}
			
The rc-summary, defined in \Cref{sec:operations}, of the concatenation of sequences $S_1$ and $S_2$ is computed by \Cref{combinerc}. 
			\begin{algorithm}[H]
			  \small
			  \caption{\small Implementation of \textsc{combine-rc-summaries}\label{combinerc}}
			  \begin{algorithmic}[1]
			    \Procedure{combine-rc-summaries}{$\var{s1},\var{s2},\oplus$}
				    \State \var{answer} $\gets$ empty rc-summary
				    \If {\var{s1.suffix-depth} + \var{s2.prefix-depth} $> 0$}
				    	\State \var{answer.prefix-depth} $\gets$ \var{s1.prefix-depth}
				    	\State \var{answer.body-combination} $\gets$ \var{s1.body-combination}
				    	\State \var{answer.suffix-depth} $\gets$ \var{s1.suffix-depth} + \var{s2.prefix-depth} + \var{s2.suffix-depth}
				    	\State \var{answer.suffix-info} $\gets$ \var{s1.suffix-info}
				    \EndIf
				    \If {\var{s1.suffix-depth} + \var{s2.prefix-depth} $< 0$}
				    	\State \var{answer.prefix-depth} $\gets$ \var{s1.prefix-depth} + \var{s1.suffix-depth} + \var{s2.prefix-depth}
				    	\State \var{answer.body-combination} $\gets$ \var{s2.body-combination}
				    	\State \var{answer.suffix-depth} $\gets$ \var{s2.suffix-depth}
				    	\State \var{answer.suffix-info} $\gets$ \var{s2.suffix-info}
				    \EndIf
				    \If {\var{s1.suffix-depth} + \var{s2.prefix-depth} $= 0$}
				    	\State \var{answer.prefix-depth} $\gets$ \var{s1.prefix-depth}
				    	\State \var{answer.body-combination} $\gets$ \var{s2.body-combination} $\oplus$ \var{s1.body-combination} $\oplus$ \var{s1.suffix-info}
				    	\State \var{answer.suffix-depth} $\gets$ \var{s2.suffix-depth}
				    	\State \var{answer.suffix-info} $\gets$ \var{s2.suffix-info}
				    \EndIf
			    \EndProcedure
			  \end{algorithmic}
			\end{algorithm}
			We can augment the splay tree nodes as before, keeping track of the summary combination for every range associated with the nodes of the splay tree. The result of \textsc{combine-children}$(v, \oplus)$ is equal to to \var{body-combination} field of the rc-summary of the range starting in the successor of \var{open-v} and ending in the predecessor of \var{close-v}.
			
			To support \textsc{combine-child-subtree} we need to extend the definition of rc-summaries to keep track of the partial combination in the prefix and the suffix.
	
For the sake of completeness we report below the summary used by \textsc{reduce-child-subtrees}. 
\begin{definition}[rcs-summary]
An \emph{rcs-summary} of a sequence of parentheses is a tuple having these fields:
\begin{compactitem}
	\item {{\var{prefix-depth}}}, the depth of the minimal-depth parenthesis
	\item {{\var{prefix-$\oplus$-info}}}, the $\oplus$-combination of the values of the nodes associated with the prefix 
	\item {{\var{body-$\oplus$-info}}}, the $\oplus$-combination of the values of the nodes associated with the body
	\item {{\var{body-$\otimes$-info}}}, the $\otimes$-combination of the $\Sigma$-values of the subtrees in the body
	\item {{\var{suffix-$\oplus$-info}}}, the $\oplus$-combination of the values of the nodes associated with the body
	\item {{\var{suffix-depth}}}, the difference between the depth of the last parenthesis and the depth of any minimal-depth parenthesis.
\end{compactitem}
\end{definition}

\subsection{Applications: betweenness and closeness centrality}

We now need to show how to maintain the information related to \textsc{up-dists} and \textsc{down-dists} when we perform the following structural updates:
	\begin{compactitem}
		\item \textsc{link}
		\item \textsc{cut}
		\item \textsc{condense}
	\end{compactitem}
	Note that the other structural updates are maintained: \textsc{evert} is implemented using \textsc{link} and  \textsc{cut}; \textsc{erase} is implemented using \textsc{cut} and  \textsc{condense}.

\noindent \textbf{\textsc{link}}. As we mentioned in \Cref{sub:cc}, in the case of a \textsc{link} operation, where we add the edge between $u$ and $v$, whose weight is $w$, the following operations need to be executed before the actual linking to maintain the information (we denote the size of the tree $u$ (resp. $v$) belongs to with $s_u$ (resp. $s_v$)):
\begin{compactitem}
\item the \textsc{down-dists} of all the nodes in the path of $u$ are increased by $w\cdot\textsc{subtree-size}(v) + \textsc{down-dists}(v)$;
\item the \textsc{up-dists} of all the nodes in the subtree of $v$ (included) are increased by $w\cdot\textsc{subtree-size}(\textsc{root}(u)) + \textsc{up-dists}(u) + \textsc{down-dists}(v)$;
\item the \textsc{up-dists} of all the nodes in the tree containing $u$, with the only exception of the nodes in the path of $u$, are increased by $w\cdot\textsc{subtree-size}(v) + \textsc{down-dists}(v)$. In order to do so, we add it to all the nodes (i.e. the subtree of \textsc{root}$(u)$), and then we subtract it from all the nodes in the path of $u$. 
\end{compactitem}

\noindent	\textbf{\textsc{cut}}. The \textsc{cut} is the dual of the \textsc{link}, thus we execute the following operations after the cut:
\begin{compactitem}
\item the \textsc{down-dists} of all the nodes in the path of $u$ are decreased by $w\cdot\textsc{subtree-size}(v) + \textsc{down-dists}(v)$;
\item the \textsc{up-dists} of all the nodes in the subtree of $v$ (included) are decreased by $w\cdot\textsc{subtree-size}(\textsc{root}(u)) + \textsc{up-dists}(u) + \textsc{down-dists}(v)$;
\item the \textsc{up-dists} of all the nodes in the tree containing $u$, with the only exception of the nodes in the path of $u$, are decreased by $w\cdot\textsc{subtree-size}(v) + \textsc{down-dists}(v)$. In order to do so, we subtract if from all the nodes (i.e. the subtree of \textsc{root}$(u)$), and then we add it to all the nodes in the path of $u$. 
\end{compactitem}
		
\noindent	\textbf{\textsc{condense}}
When we condense node $v$, let us denote by $u$ the parent of $v$ and by $w$ the weight of the edge ($u$,$v$). We execute the following operations before condensing:
\begin{compactitem}
\item the \textsc{down-dists} of all the nodes in the path of $u$ are decreased by $w\cdot\textsc{subtree-size}(v)$;
\item the \textsc{up-dists} of all the nodes in the subtree of $v$ (included) are decreased by $w\cdot(\textsc{subtree-size}(\textsc{root}(v)) - \textsc{subtree-size}(v))$;
\item the \textsc{up-dists} of all the nodes in the tree containing $u$, with the only exception of the nodes in the path of $u$, are decreased by $w\cdot\textsc{subtree-size}(v)$. In order to do so, we subtract if from all the nodes (i.e. the subtree of \textsc{root}$(u)$), and then we add it to all the nodes in the path of $u$. 
\end{compactitem}

We now detail how to maintain a value in the node, such as \textsc{down-dists} and \textsc{up-dists}, under the two following operations: \textsc{add-to-path}$(v,\delta)$ that adds $\delta$ to all the vertices in the path between $v$ and the root, and \textsc{add-to-subtree}$(v,\delta)$ that adds $\delta$ to all the vertices in the subtree of $v$. 
In each node we 	maintain the following information, that will be used to derive the value of the node\footnote{Thus, in order to maintain both \textsc{down-dists} and \textsc{up-dists} we need six distinct values in a node: a $\Delta_\uparrow, \Delta_\downarrow,$ and  $\Delta_\bullet$ for \textsc{down-dists}, and a $\Delta_\uparrow, \Delta_\downarrow,$ and  $\Delta_\bullet$ for \textsc{up-dists}.}:
\begin{compactitem}
				\item $\Delta_\uparrow$, to be forwarded in the path of the node;
				\item $\Delta_\downarrow$, to be forwarded in the subtree of the node;
				\item $\Delta_\bullet$, relative to the node.
			\end{compactitem}

In the begininning $\Delta_\uparrow$ and $\Delta_\downarrow$ are equal to $0$, whilst $\Delta_\bullet$ has the initial value of the node.

This allow us to state the following Lemma.
\begin{restatable}{lemma}{bc}
Using a \dft, it is possible to answer \emph{closeness centrality} queries of a vertex in time $\bigoh (\log n)$.
\end{restatable}

In the following we report the pseudocode of the affected operations, where we show the changes from the previously shown pseudocodes in red (best viewed in color).

			\begin{algorithm}[H]
			  \small
			  \caption{\small Implementation of \textsc{get-effective-val}}
			  \begin{algorithmic}[1]
			    \Procedure{get-effective-val}{$v$}\Comment{$\bigoh(\log n)$}
			    	\State \textbf{return} $\Delta_\bullet(v)$ + (sum of $\Delta_\uparrow(v)$ in the subtree of $v$) + (sum of $\Delta_\downarrow(v)$ in the path of $v$)
			    \EndProcedure
			  \end{algorithmic}
			\end{algorithm}
			
			\begin{algorithm}[H]
			  \small
			  \caption{\small Implementation of \textsc{increment-val} -- increase the value of node $v$ by $\delta$}
			  \begin{algorithmic}[1]
			    \Procedure{increment-val}{$v, \delta$}\Comment{$\bigoh(1)$}
			    	\State $\Delta_\bullet(v) \gets \Delta_\bullet(v) + \delta$
			    \EndProcedure
			  \end{algorithmic}
			\end{algorithm}

			\begin{algorithm}[H]
			  \small
			  \caption{\small Implementation of \textsc{change-val} -- set the value of node $v$ to $\var{target}$}
			  \begin{algorithmic}[1]
			    \Procedure{change-val}{$v, \var{target}$}\Comment{$\bigoh(\log n)$}
			    	\State $\delta \gets \var{target} - \textsc{get-effective-val}(v)$
			    	\State $\textsc{increment-val}(v, \delta)$
			    \EndProcedure
			  \end{algorithmic}
			\end{algorithm}

			\begin{algorithm}[H]
			  \small
			  \caption{\small Implementation of \textsc{link}}
			  \begin{algorithmic}[1]
			    \Procedure{link}{$u,v$}
			    \If{\textbf{not} \textsc{same-tree}($u, v$)}
			    	\State\textcolor{red}{$\Delta_\uparrow(u) \gets \Delta_\uparrow(u)$ - sum of $\Delta_\uparrow$ of subtree of $v$}
			    	\State\textcolor{red}{$\Delta_\downarrow(v) \gets \Delta_\downarrow(v)$ - sum of $\Delta_\downarrow$ in the path of $u$}
				    \State \var{open-u} $\gets$ open-node of node $u$ in the dft
				   	\State \var{close-u} $\gets$ close-node of node $u$ in the dft
				   	\State \var{open-v} $\gets$ open-node of node $v$ in the dft
			    	\State \textsc{splay-split}(\var{open-u})
			    	\State \textsc{splay-merge}(\var{open-u}, \var{open-v})
			    	\State \textsc{splay-merge}(\var{open-u}, \var{close-u})
			    \EndIf
			    \EndProcedure
			  \end{algorithmic}
			\end{algorithm}	
			
			\begin{algorithm}[H]
			  \small
			  \caption{\small Implementation of \textsc{cut}}
			  \begin{algorithmic}[1]
			    \Procedure{cut}{$v$}
			    \State \var{root} $\gets$ \textsc{root}($v$)
			    \If{$v$ $\neq$ \var{root}}
			    	\State\textcolor{red}{$\Delta_\uparrow(\textsc{parent}(v)) \gets \Delta_\uparrow(\textsc{parent}(v))$ + sum $\Delta_\uparrow$ in the subtree of $v$}
			    	\State\textcolor{red}{$\Delta_\downarrow(v) \gets \Delta_\downarrow(v)$ + sum of  $\Delta_\downarrow$ in the path of $\textsc{parent}(v)$}
					\State \var{open-v} $\gets$ open-node of node $v$ in the dft
					\State \var{close-v} $\gets$ close-node of node $v$ in the dft
					\State \var{open-root} $\gets$ open-node of \var{root} in the dft
					\State \var{close-root} $\gets$ close-node of \var{root} in the dft
					\State \textsc{splay-split}(\textsc{splay-predecessor}(\var{open-v}))
					\State \textsc{splay-split}(\var{close-v})
					\State \textsc{splay-merge}(\var{open-root}, \var{close-root})
			    \EndIf
			    \EndProcedure
			  \end{algorithmic}
			\end{algorithm}
			
			\begin{algorithm}[H]
			  \small
			  \caption{\small Implementation of \textsc{condense}}
			  \begin{algorithmic}[1]
			    \Procedure{condense}{$v$}
			    \If{\textcolor{red}{$v \neq \textsc{root}(v)$}}
					\State\textcolor{red}{$\Delta_\uparrow(\textsc{parent}(v))$ = $\Delta_\uparrow(\textsc{parent}(v)) + \Delta_\uparrow(v)$}
					\State\textcolor{red}{$\Delta_\downarrow(\textsc{parent}(v))$ = $\Delta_\downarrow(\textsc{parent}(v)) + \Delta_\downarrow(v)$}
					\State\textcolor{red}{$\Delta_\bullet(\textsc{parent}(v))$ = $\Delta_\bullet(\textsc{parent}(v)) - \Delta_\downarrow(v)$}
				\EndIf
				\State \var{open-v} $\gets$ open-node of node $v$ in the dft
				\State \var{close-v} $\gets$ close-node of node $v$ in the dft
				\State \textsc{splay-erase}(\var{open-v})
				\State \textsc{splay-erase}(\var{close-v})
			    \EndProcedure
			  \end{algorithmic}
			\end{algorithm}

\end{document}